\documentclass[a4paper,fleqn]{cas-sc}
\usepackage{todonotes}
\usepackage[numbers]{natbib}
\usepackage{subcaption}
\def\tsc#1{\csdef{#1}{\textsc{\lowercase{#1}}\xspace}}
\tsc{WGM}
\tsc{QE}
\begin{document}
\let\WriteBookmarks\relax
\def\floatpagepagefraction{1}
\def\textpagefraction{.001}

\shorttitle{Impact of Atrial Fibrillation on Left Atrium Haemodynamics}    

\shortauthors{M. Corti, A. Zingaro, L. Dede', A. Quarteroni}  

\title [mode = title]{Impact of Atrial Fibrillation on Left Atrium Haemodynamics: A Computational Fluid Dynamics Study}  

%\tnotemark[<tnote number>] 

%\tnotetext[<tnote number>]{<tnote text>} 


\author[Polimi]{Mattia Corti}[orcid=0000-0002-7014-972X]

\cormark[1]


\ead{mattia.corti@polimi.it}

%\ead[url]{<URL>}

%\credit{<Credit authorship details>}

\affiliation[Polimi]
{organization={MOX-Dipartimento di Matematica, Politecnico di Milano},%Department and Organization
            addressline={Piazza Leonardo da Vinci 32}, 
            city={Milan},
            postcode={20133},
            country={Italy}}
            
\author[Polimi]{Alberto Zingaro}
%\ead{alberto.zingaro@polimi.it}
\author[Polimi]{Luca Dede'}
%\ead{luca.dede@polimi.it}
\author[Polimi,EPFL]{Alfio Maria Quarteroni}
%\ead{alfio.quarteroni@polimi.it}

\affiliation[EPFL]{organization={Institute of Mathematics, \'{E}cole Polytechnique F\'{e}d\'{e}rale de Lausanne},%Department and Organization
            addressline={Station 8, Av. Piccard}, 
            city={Lausanne},
            postcode={CH-1015}, 
            country={Switzerland (Professor Emeritus)}}


\begin{abstract}
We analyse the haemodynamics of the left atrium, highlighting differences between healthy individuals and patients affected by atrial fibrillation. The computational study is based on patient-specific geometries of the left atria to simulate blood flow dynamics. We design a novel procedure to compute the boundary data for the 3D haemodynamic simulations, which are particularly useful in absence of data from clinical measurements. With this aim, we introduce a parametric definition of atrial displacement, and we use a closed-loop lumped parameter model of the whole cardiovascular circulation conveniently tuned on the basis of the patient's characteristics. We evaluate several fluid dynamics indicators for atrial haemodynamics, validating our numerical results in terms of clinical measurements; we investigate the impact of geometric and clinical characteristics on the risk of thrombosis. To highlight the correlation of thrombus formation with atrial fibrillation, according to medical evidence, we propose a novel indicator: age stasis. It arises from the combination of Eulerian and Lagrangian quantities. This indicator identifies regions where slow flow cannot properly rinse the chamber, accumulating stale blood particles, and creating optimal conditions for clots formation.
\end{abstract}


%\item CFD atrial simulations with BCs from a 0D model with patient-specific constraints.
%\item 0D circulation model and parametric displacement definition accounting for AF.
%\item A novel haemodynamic indicator combining Eulerian and Lagrangian perspectives.
%\item LAA morphological features strongly impact the haemodynamics affecting its washout.
%\item Characterisation of haemodynamics in AF conditions with different levels of severity.

% Keywords

\maketitle

\section{Introduction}
Atrial fibrillation (AF) is the most common cardiac electric dysfunction worldwide \cite{AF:dicarlo}. The irregular electrical impulses of this pathology cause a reduced atrial contraction and thus a smaller blood ejection. According to the European Society of Cardiology (ESC), in 2016, 7.6 million people aged 65 and over were affected by AF in the European Union. Figures would increase up to 14.4 million within 2060 \cite{AF:dicarlo}.

\par

In terms of pathology severity, AF is divided into three categories: paroxysmal AF is an episode that typically self-terminates within seven days; persistent AF requires termination by pharmacological or direct-current electric cardioversion; permanent AF is irreversible to sinus rhythm \cite{iwasaki:AF,kowey:AF,schotten:AF}. Persistent AF can cause long-term remodeling of the atrial chambers, increasing atrial volume and causing thrombogenic formation in the Left Atrial Appendage (LAA) \cite{azzam:inbook,sanfilippo:AF}. Two centuries ago, ``Virchow's triad'' was defined to denote the three main factors contributing to the risk of thrombosis: endothelial injuries, hypercoagulability, and blood stasis \cite{virchow:triad}. The correlation between these elements and AF is nowadays established \cite{azzam:inbook,virchow:rev}.

\par

In this paper, we investigate the effects of AF on instantaneous cardiac haemodynamics. Cardiac blood flow analysis is commonly assessed using both imaging and experimental techniques. For example, 4D flow MRI \cite{4DMRI:ventricles}, one of the most advanced imaging techniques, allows the detection of a time-dependent blood flow fields \cite{4DMRI:markl}, the estimation of haemodynamic parameters such as flow stasis, mean velocity \cite{4DMRI:stasis}, and particle tracking. However, the resolution provided by 4D flow MRI might not be enough to accurately catch the complexity of cardiac flows and their transitional effects: the formation of shear layers, small vortices, and their interactions \cite{CFD4D:cerebral, CFD4D:mix, CFD4D:markl, burriesci:MRI, roldan2015hemodynamic}. For this reason, in-silico simulations of the heart, often combined with medical images, stand as a valuable tool for a more accurate description of blood flows, using haemodynamic indicators as the wall shear stress (WSS) \cite{chnafa:LES, fedele:Mitral, pamplona:age, karabelas:4chambers}. 

\par

Literature abounds with CFD studies of human atria under AF, both for idealized  \cite{zhang:2008, burriesci:2} and patient-specific geometries \cite{koizumi:2014, otani:2016, masci:2017, masci:2017bis, thermal:CFD, garcia:1, mill:1, DUENASPAMPLONA202227}. Concerning the numerical approach, in \cite{otani:2016}, CFD simulations were performed without the application of a turbulence model; in \cite{masci:2017, masci:2017bis}, the LA haemodynamics is modelled via the Navier-Stokes (NS) equations in Arbitrary Lagrangian-Eulerian (ALE) formulation; the Variational Multiscale Large Eddy Simulation (VMS-LES) method \cite{forti:BDF} is used to account for possible transitional-to-turbulent flows. Moreover, a comparison is made considering the differences derived from applying or not an LES model in AF conditions in \cite{pamplona:age}. They numerically demonstrated that the absence of a turbulence model is acceptable in AF conditions.
\par
Due to the relevance of LAA in thrombus formation, many studies investigate how the geometrical morphology of this region of the LA affects blood flow \cite{masci:2019, WSS:garcia, bosi:2018, CFD:WSS, ECAP:LAA2, dongjie:2019, pons:1}. These works suggest the existence of a strong correlation between LAA morphology and thromboembolic risk; moreover, AF aggravates this danger. 

\par

In this paper, we consider patient-specific geometries of the LA and we carry out CFD simulations that provide a complete characterisation of blood flow under physiological and pathological conditions. In particular, the atrial geometries we have available \cite{niederer:geom, roney:geom} are scanned at the end of diastole only \cite{pash:geom}. Thus, we cannot derive any clinical information in terms of boundary pressures, flowrates, and displacement. \textcolor{black}{Thus, a 0D closed-loop circulation model \cite{reg:primo}  serves as input to prescribe boundary conditions to the 3D CFD problem, by employing a ``one-way'' 3D-0D coupling scheme. More precisely, we customize the closed-loop circulation model with the available patient-specific data to get transient data that we prescribe on the boundary of the CFD domain. Specifically, to simulate AF conditions, we conveniently tune the circulation model as explained in \cite{scarsoglio:AF}. Thus, the proposed procedure allows to carry out numerical simulations also when time dependent patient-specific data are not available. Moreover, since fluxes and displacement are coming from the same circulation model, the mass conservation property of the 0D closed-loop model is naturally encoded in the 3D boundary conditions. This guarantees to satisfy the compatibility condition of the NS equations \cite{quartapelle:CFD} that are required for the well-posedness of the problem. On the contrary, fluxes and displacements obtained from measurements not related to the same patient would not guarantee this property, thus affecting the meaningfulness of the corresponding numerical solution.}

\par

Contextually, we tune additional model parameters to account for the volumetric constraints given by the patient-specific atrial geometries. We also used this calibration process to obtain meaningful Left Atrial Ejection Fractions (LAEF). Similarly to \cite{zingaro:LA, dede:leftheart}, we use a parametric displacement combined with patient-specific geometries to fill the lack of information on the chamber displacement. Starting from the analytical formulation in \cite{zingaro:LA}, that prescribed a movement directed towards the centre of mass of LA, we modify it to capture a more realistic displacement of the LAA and to match the volume variations with the physiological or pathological values of the Left Atrial Appendage Ejection Fraction (LAAEF) \cite{LAAEF:gan}. We generate a displacement field that embodies patient-specific constraints to obtain a more physiological motion of the atrial chamber and of its auricle. We believe that this is essential to estimate the risk of thrombosis in the LA, considering that LAA is the area where thrombi formation begins \cite{azzam:inbook}. Furthermore, a parametric displacement allows us to simulate different levels of severity of AF according to the clinical situation of the patients by conveniently changing the parameters involved in the displacement definition. \textcolor{black}{However, this model relies on a number of assumptions that we cannot entirely validate by carrying out a direct comparison between our displacement and some in-vivo recordings, since kinematic data are not available. Thus,  in order to assess the correctness of our numerical results, we compare a numbers of in-silico values with biomarkers available in literature that are acquired in healthy and pathological patients. We show that the computed values always lie in the given ranges. Furthermore, we highlight that the methodology we employ aims to carry out LA haemodynamic simulations in both physiologic and AF conditions, overcoming the problem of missing data, and with a contained computational cost.}

\par

In many works on atrial haemodynamic simulations, the effects of the mitral valve (MV) are mimicked by employing switching boundary conditions \cite{masci:2017, masci:2017bis, burriesci:2, otani:2016}. Differently, in this paper, we model the effect of the MV on the fluid flow through the Resistive Immersed Implicit Surface (RIIS) method \cite{fedele:RIIS, fedele:Mitral, zingaro2022modeling}. Furthermore, we prescribe valvular opening and closing times that are consistent with clinical findings, overcoming the classical oversimplification of an instantaneous switch of the valvular status \cite{masci:2017, masci:2017bis, burriesci:2, otani:2016}. To the best of our knowledge, the only work in the literature that simulates left atrial haemodynamics considering the presence of MV is \cite{niederer:MV}, where a fluid-structure interaction model is employed to perform an advanced analysis of the valvular motion.

\par

\textcolor{black}{Finally, we analyze some haemodynamic indicators both from Eulerian and Lagrangian perspectives. The Eulerian indicators are directly derived from the results of the NS simulations. Differently, the Lagrangian ones are obtained by simulating the red blood cells motion in the atrial chamber.  We consider particles like tracers (their presence does not influence the blood flow) \cite{chnafa:LES}, and, taking advantage of the kinetic theory development for particles transport \cite{zaichik:particle}, we derive some Lagrangian fields as mean age and washout of the blood \cite{sierra:age}. The combination of these two approaches has been successfully applied in the literature to analyse blood flow in ventricles \cite{chnafa:LES, Rossini2016ACM}. In this paper, for the first time, we propose a new hybrid indicator, that we call \textit{age stasis}, which is defined as the product between a Lagrangian term and an Eulerian one. It detects regions with high thrombotic risk, where stagnant flow and high blood mean age subsist at the same time.} As a matter of fact, the coexistence of these two situations denotes a high risk of blood clots formation. Furthermore, we derive a dimensionless indicator that estimates the percentage of volume associated with a higher risk and exploring its correlation with the AF pathology. This allows to carry out comparisons among different patients in terms of a single, synthetic indicator.

\par
\textcolor{black}{This paper is organised as follows: we present the mathematical models and methods in Section~\ref{sec:methods}. Finally, in Section~\ref{sec:results}, we present the numerical results of both the $0D-$model and CFD simulation. In particular, in Section~\ref{sec:agestasis}, we propose our haemodynamic indicator. The discussion of the results is reported in Section~\ref{sec:disc}. Eventually, conclusions are drawn in Section~\ref{sec:conclusions}.}

\section{Methods}
\label{sec:methods}
In this section, we introduce the methods we employ to carry out CFD simulations. Particularly, in Section~\ref{sec:NS}, we introduce the mathematical models and numerical methods.  Section~\ref{sec:BCs} is devoted to the description of boundary conditions obtained via the lumped-parameter circulation model and the parametrization of the atrial displacement. Section \ref{sec:NSS} concerns the setup of the CFD simulations.
\color{black}
\subsection{Mathematical models and numerical methods for left atrial haemodynamics}
\label{sec:NS}
This section introduces the mathematical models to describe the fluid dynamics in LA. To carry out the simulations, we use the medical images that were obtained by \cite{pash:geom}. The derived endocardial geometries are openly accessible from the supplementary material of \cite{niederer:geom, roney:geom}. Data are scanned at the time of end diastole only. Thus, we do not have any knowledge in terms of displacement field, boundary flowrates and boundary pressures. 
%However, the construction of boundary conditions uses the end diastolic volume of both left and right atria derived from the images.
\par
\begin{figure}[t]
	\centering
	{\includegraphics[width=0.7\textwidth]{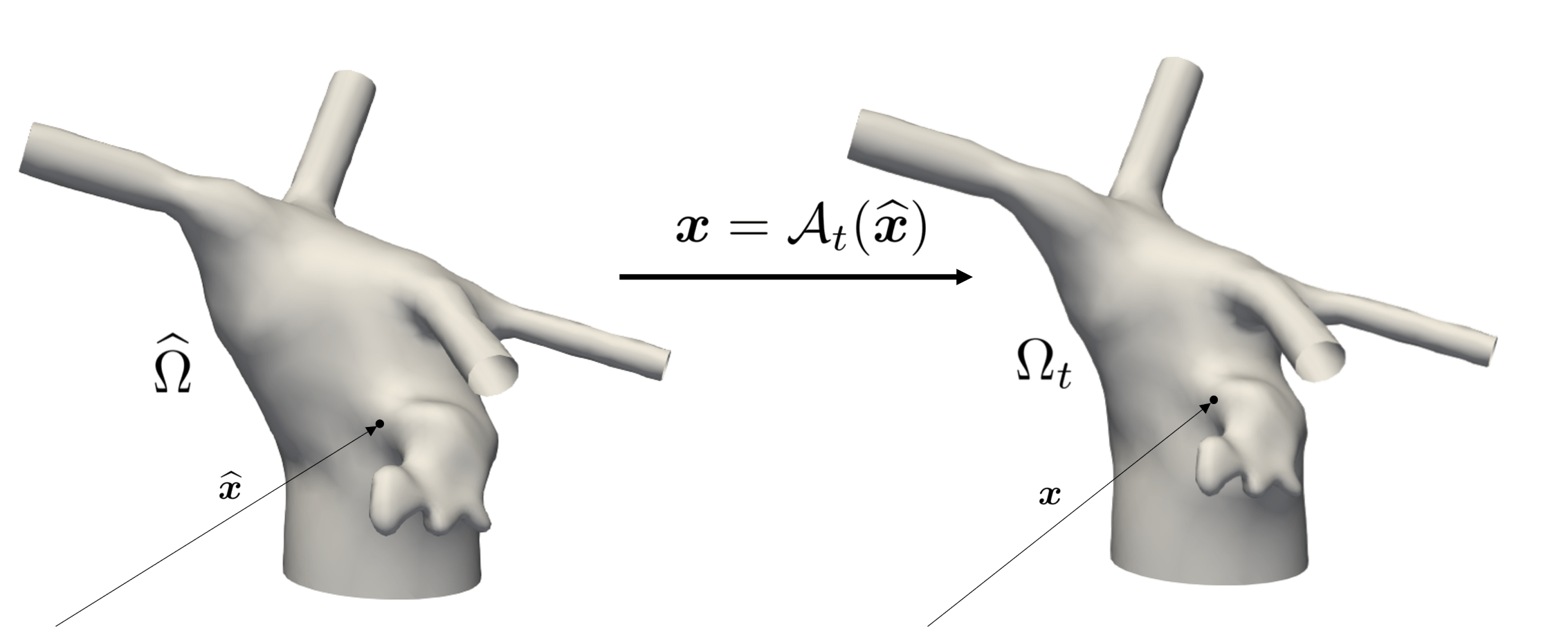}}
	\caption{\textcolor{black}{Left atrial domain in reference configuration $\widehat \Omega$ (left), in current configuration $\Omega_t$ (right), and the ALE map $\boldsymbol{x}=\boldsymbol{\mathcal{A}}_t(\widehat{\boldsymbol{x}})$.}}
	\label{fig:ALEdomain}
\end{figure}

Let $\Omega_t$ be the fluid domain at a specific time instant $t\in(0,T)$ (current configuration) and let $\partial \Omega_t$ be its boundary, being $T$ the final time. To take into account the moving reference framework, we employ the ALE formulation \cite{donea:ALE}. Let  $\widehat{\Omega}\subset\mathbb{R}^3$ be the LA domain in its reference configuration, as displayed in Figure~\ref{fig:ALEdomain}. We define the ALE map $\boldsymbol{\mathcal{A}}_t$, which associates at each point of the reference configuration $\widehat{\boldsymbol{x}}$ the corresponding point in the actual one $\boldsymbol{x}$, such that $
\boldsymbol{\mathcal{A}}_t:\widehat{\Omega}\rightarrow\Omega_t$ and $\boldsymbol{x}=\boldsymbol{\mathcal{A}}_t(\widehat{\boldsymbol{x}}) = \widehat{\boldsymbol{x}} + \widehat{\boldsymbol{d}}(\widehat{\boldsymbol{x}},t)$,
being $\widehat{\boldsymbol{d}}$ the displacement with respect to the reference configuration.
\par
\begin{figure}[t]
	\centering
	{\includegraphics[width=0.7\textwidth]{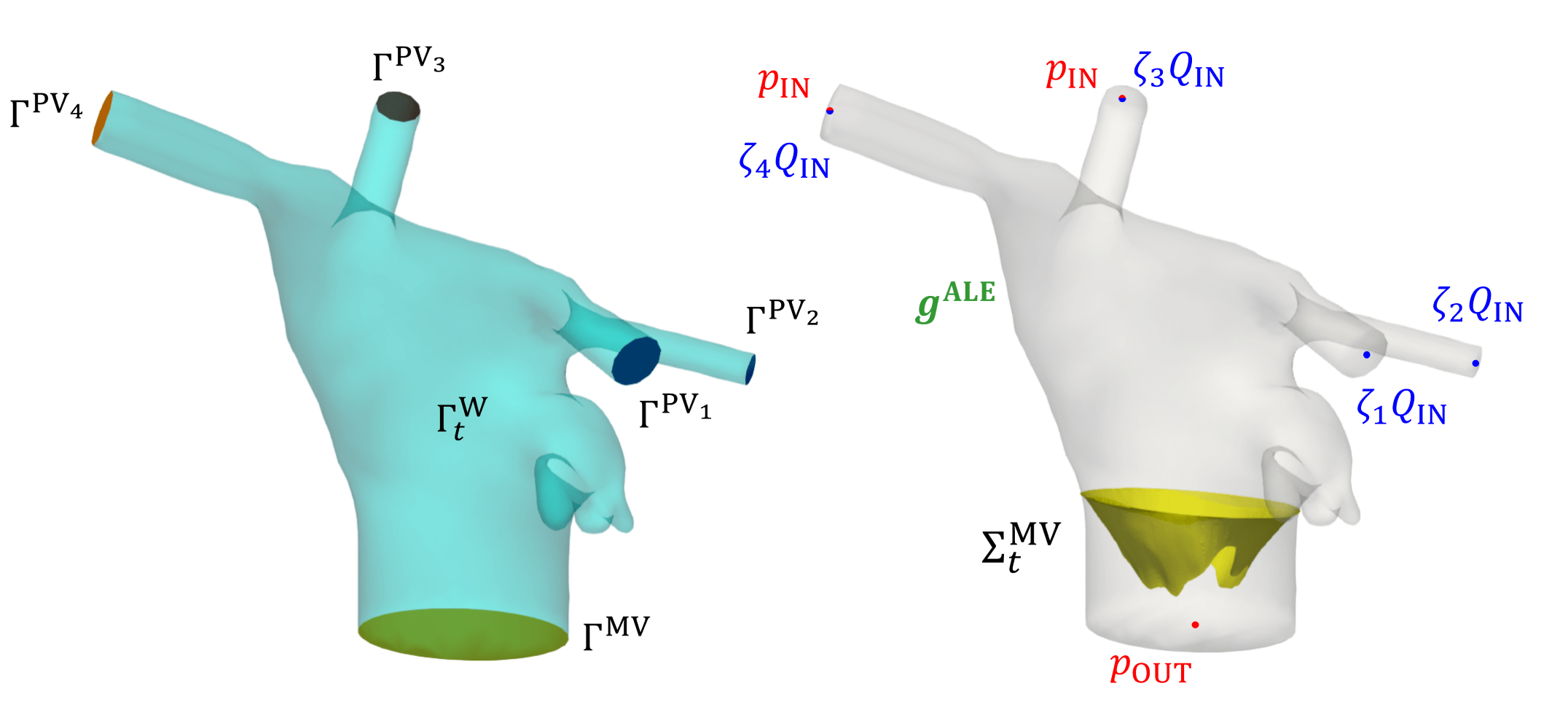}}
	\caption{LA domain with details on the boundaries (left), and with the representation of the MV surface $\Sigma^\mathrm{MV}_t$ and of the boundary data that we prescribe (right).}
	\label{fig:DomainBCs}
\end{figure}

As shown in Figure~\ref{fig:DomainBCs}, we split the boundary as $\partial\Omega_t = \Gamma_t^\mathrm{W} \cup \Gamma^\mathrm{MV} \cup \Big( \bigcup_{j=1}^{4} \Gamma^{\mathrm{PV}_j} \Big)$, being $\Gamma_t^\mathrm{W}$ the endocardial wall, $\Gamma^{\mathrm{PV}_j}$ the $j$-th pulmonary vein inlet section, with $ j = 1,\dots,4$, and $\Gamma^\mathrm{MV}$ the outlet section downstream of the MV. For the sake of simplicity, we consider the inlet and outlet sections to be fixed, neglecting hence the time dependency in the notation. 
\par
Should we know a boundary velocity $\boldsymbol{g}^\mathrm{ALE}:\partial\Omega_t\times(0,T)\rightarrow \mathbb{R}^3$, we recover the ALE velocity $\boldsymbol{u}^\mathrm{ALE}:\Omega_t\times(0,T)\rightarrow \mathbb{R}^3$ by means of the following harmonic extension problem:
\begin{equation}
	\begin{cases}
		-\Delta\boldsymbol{u}^\mathrm{ALE}=\boldsymbol{0} & \mathrm{in}\;\Omega_t \times (0, T), \\
		\boldsymbol{u}^\mathrm{ALE}=\boldsymbol{g}^\mathrm{ALE} & \mathrm{on}\;\partial\Omega_t \times (0, T), \\
	\end{cases}
	\label{eq:harm}
\end{equation}
which allows regaining displacement as: $ \boldsymbol{d}(\boldsymbol{x},t) = \int_0^t \boldsymbol{u}^\mathrm{ALE}(\boldsymbol{x},\tau)d\tau$. 
\par
In the heart chambers, it is assumed that blood behaves as a Newtonian, incompressible, and homogeneous fluid \cite{vergara:actanumerica, blood:newton}. Under this assumption, the Cauchy's stress tensor is defined as $\boldsymbol{\sigma}(\boldsymbol{u},p) = - p \mathbf{I} + \mu\left(\nabla \boldsymbol{u} + (\nabla \boldsymbol{u})^\top \right)$,
where $\boldsymbol{u}:\Omega_t\times(0,T)\rightarrow\mathbb{R}^3$ is the fluid velocity field, $p:\Omega_t\times(0,T)\rightarrow\mathbb{R}$ the pressure field and $\mu$ the dynamic viscosity. Then, the fluid dynamics equations read:
\begin{equation}
\label{eq:NS-nobc}
	\begin{cases}
		\rho\left(\dfrac{\widehat{\partial}\boldsymbol{u}}{\widehat{\partial} t} + ((\boldsymbol{u}-\boldsymbol{u}^\mathrm{ALE})\cdot\nabla)\boldsymbol{u}\right)-\nabla\cdot\boldsymbol{\sigma}(\boldsymbol{u},p) + \dfrac{R}{\varepsilon}\delta_\varepsilon^\Sigma(\varphi)(\boldsymbol{u}-\boldsymbol{u}_\Sigma) = \boldsymbol{0} & \mathrm{in}\;\Omega_t\times(0,T),\\
		\nabla\cdot\boldsymbol{u}=0 & \mathrm{in}\;\Omega_t\times(0,T).
	\end{cases}
\end{equation}
\par
To account for the presence of the MV in the fluid, in Equation \ref{eq:NS-nobc}, we use the Resistive Immersed Implicit Surface (RIIS) method, proposed by \cite{fedele:RIIS} for the simulation of the aortic valve, and extended to the ALE case in \cite{fedele:Mitral}. With the RIIS method, we identify the MV as an immersed surface described by the level set function $\varphi:\Omega_t\rightarrow\mathbb{R}$ as $\Sigma^\mathrm{MV}_t = \{ \boldsymbol{x}\in\Omega_t: \; \varphi(\boldsymbol{x}) = 0\}$. Moreover, $\varepsilon$ is a parameter representing the half-thickness of the MV leaflets, $R$ the valve resistance, and $\delta^\Sigma_\varepsilon$ is the smooth Dirac delta function defined as in \cite{fedele:RIIS}. % follows:
%\begin{equation}
%	\delta_\varepsilon^\Sigma(\varphi) = 
%	\begin{cases}
%		\dfrac{1}{2\varepsilon}\Big(1+\cos\Big(\dfrac{\pi\varphi}{\varepsilon}\Big)\Big) & \mathrm{if}\,|\varphi|\leq\varepsilon, \\
%		0 	& \mathrm{if}|\varphi|\,>\varepsilon.
%	\end{cases}
%\end{equation}
The valve velocity is set to be null, using the quasi-static approximation $(\boldsymbol{u}_\Sigma = \boldsymbol{0})$ \cite{fedele:Mitral, zingaro2022modeling}.
\subsubsection{Boundary and initial conditions}
We apply a nonhomogeneous Dirichlet no-slip condition $\boldsymbol{u}=\boldsymbol{g}^\mathrm{ALE}$ on the endocardial wall $\Gamma_t^\mathrm{W}$; the boundary datum $\boldsymbol{g}^\mathrm{ALE}$ is obtained as presented in Section \ref{sec:BCs}. On the MV section $\Gamma^\mathrm{MV}$, we impose an outflow Neumann boundary condition considering as mean stress value $p_\mathrm{OUT}$. We consider heartbeat and diastole duration $T_\mathrm{HB}$ and $T_\mathrm{d}$, respectively.
\par
Regarding the inlet sections $\Gamma^\mathrm{PV}_j$, we use a Dirichlet boundary condition for all veins in the diastolic phase $(0,T_\mathrm{d}]$, imposing an inlet flux $Q_\mathrm{IN}$. However, in principle, this would be a defective condition, since it prescribes only one scalar function through the section and not the overall velocity field \cite{vergara:actanumerica}. A possible way to fill this gap is to prescribe a parabolic velocity profile to complete the information. In the systolic phase of the heartbeat $(T_\mathrm{d},T_\mathrm{HB}]$, the closed MV would not allow a correct estimate of the atrial pressure without imposition of a Neumann condition on some veins. For this reason, we switch the boundary conditions in two inlet sections, by prescribing the mean pressure value $p_\mathrm{IN}$, as done in \cite{vergara:actanumerica}. For the switching BCs, following arguments of \cite{murraylaw}, we set the same pressures in vessels of the same size. Thus, we choose the two veins being characterized by the most similar cross-section areas. Moreover, to (weakly) penalise the reverse flow, we introduce backflow stabilization in all the Neumann boundaries to avoid numerical instabilities \cite{bertoglio:backflow}. 

We denote by $\tilde{R}_j$ the radius of the $j-$th inlet section; $r(\boldsymbol{x})=|\boldsymbol{x}|_2$ is the radial coordinate of the point $\boldsymbol{x}\in\Gamma^{\mathrm{PV}_j}$, $|\cdot|_2:\mathbb{R}^3\rightarrow\mathbb{R}$ being the Euclidean norm. To distribute the inlet flow in veins having different cross sections, we introduce a flow repartition factor associated with the $j$--th vein. We compute it proportionally to the inlet area as:
\begin{equation}
    \label{eq:repartition}
    \zeta_j = \dfrac{|\Gamma^{\mathrm{PV}_j}|}{\sum_{k=1}^4|\Gamma^{\mathrm{PV}_k}|}.
\end{equation}
The way we compute the boundary conditions $Q_\mathrm{IN}$, $p_\mathrm{IN}$, $p_\mathrm{OUT}$ and $\boldsymbol{g}^\mathrm{ALE}$ is discussed in Section \ref{sec:BCs}.  
\par
Moreover, we consider a null initial condition  $\boldsymbol{u}(\boldsymbol{x},0) = \boldsymbol{0}$. The NS-ALE-RIIS equations with the boundary and initial conditions to simulate the LA haemodynamics read:
\par
\medskip
for every $t>0$, find $\boldsymbol{u}:\Omega_t\times(0,T)\rightarrow \mathbb{R}^3$ and $p:\Omega_t\times(0,T)\rightarrow \mathbb{R}$:
\begin{equation}
\label{eq:NS}
	\begin{cases}
		\rho\left(\dfrac{\widehat{\partial}\boldsymbol{u}}{\widehat{\partial} t} + ((\boldsymbol{u}-\boldsymbol{u}^\mathrm{ALE})\cdot\nabla)\boldsymbol{u}\right)-\nabla\cdot\boldsymbol{\sigma}(\boldsymbol{u},p) + \dfrac{R}{\varepsilon}\delta_\varepsilon^\Sigma(\varphi) \boldsymbol{u} = \boldsymbol{0} & \mathrm{in}\;\Omega_t\times(0,T_\mathrm{HB}],\\
		\nabla\cdot\boldsymbol{u}=0 & \mathrm{in}\;\Omega_t\times(0,T_\mathrm{HB}],\\
		\boldsymbol{u} = \boldsymbol{g}^\mathrm{ALE} & \mathrm{on}\;\Gamma^\mathrm{W}_t \times (0,T_\mathrm{HB}],
		\\
		\boldsymbol{u} = -2 \zeta_k \dfrac{Q_\mathrm{IN}}{|\Gamma^{\mathrm{PV}_k}|}\Big(1-\dfrac{r^2}{\tilde{R}_k^2}\Big)\boldsymbol{n}_k &  \mathrm{on}\;\Gamma^{\mathrm{PV}_k} \times (0,T_\mathrm{HB}],
		\\
		\boldsymbol{u} = -2 \zeta_j \dfrac{Q_\mathrm{IN}}{|\Gamma^{\mathrm{PV}_j}|}\Big(1-\dfrac{r^2}{\tilde{R}_j^2}\Big)\boldsymbol{n}_j &  \mathrm{on}\;\Gamma^{\mathrm{PV}_j} \times (0,T_\mathrm{d}],
		\\	\boldsymbol{\sigma}(\boldsymbol{u},p)\boldsymbol{n}_j = - p_\mathrm{IN}\boldsymbol{n}_{j} + \rho[(\boldsymbol{u}-\boldsymbol{u}^\mathrm{ALE})\cdot\boldsymbol{n}_j]_{-}(\boldsymbol{u}-\boldsymbol{u}^\mathrm{ALE}) & \mathrm{on}\;\Gamma^{\mathrm{PV}_j}\times (T_\mathrm{d},T_\mathrm{HB}],
		\\	\boldsymbol{\sigma}(\boldsymbol{u},p)\boldsymbol{n} = - p_\mathrm{OUT}\boldsymbol{n} + \rho[(\boldsymbol{u}-\boldsymbol{u}^\mathrm{ALE})\cdot\boldsymbol{n}]_{-}(\boldsymbol{u}-\boldsymbol{u}^\mathrm{ALE}) & \mathrm{on}\;\Gamma^\mathrm{MV}\times (0,T_\mathrm{HB}],
		\\
		\boldsymbol{u} = \boldsymbol{0} & \mathrm{in}\;\Omega_0\times\{0\},
	\end{cases}
\end{equation}
where $k = 1,2$ and $j = 3,4$, and $\boldsymbol{n}$ and $\boldsymbol{n}_j$ are the outgoing normals of sections $\Gamma^\mathrm{MV}$ and $\Gamma^{\mathrm{PV}_j}$, respectively. 
\subsubsection{Space and time discretizations}
Concerning the numerical approximation of Equation \ref{eq:NS}, we employ the Finite Element (FE) method for spatial discretization. We use the VMS-LES method \cite{bazilevs:VMS,forti:BDF} to obtain a stable formulation of the NS equations discretised with FE (\textit{inf-sup} condition), to stabilise the advection-dominated regime, and to account for the transitional-nearly turbulent flow according to the LES paradigm \cite{karabelas:4chambers, zingaro:LA,dede:leftheart}. \textcolor{black}{As discussed in the literature, the usage of LES methods \cite{chnafa:LES, pamplona:age}, such as the VMS-LES, become significant in cardiac applications even in presence of a transitional flow regime \cite{zingaro:LA}.} Concerning time discretization, we partition the time domain into $N_t$ time steps of equal size $\Delta t$, and we use the Backward Differentiation Formula (BDF) method of order 1. The treatment of nonlinear terms is semi-implicit with an extrapolation of the velocity field by means of the Newton-Gregory backward polynomials of first order. For more details on this method, refer to \cite{forti:BDF}. The extension of the VMS-LES method for the NS-ALE-RIIS equations can be found in \cite{zingaro:multiscale}. Analogously, a FE discretization is used to solve the lifting problem in Equation \ref{eq:harm} at each time step.
\subsection{Boundary conditions depending on circulation}
\label{sec:BCs}
Since we do not have dynamic data, but only static acquisitions of the atria at the end of diastole, we cannot recover the atrial displacement, nor the pressures and fluxes to be prescribed at the boundary. For this reason, we propose a computational procedure aimed at finding these missing data starting from a 0D circulation model and a parametric definition of the boundary displacement. This is a general procedure that can be employed when the data required to perform CFD simulations are not completely available. As we show in Section \ref{sec:NSS}, by means of this procedure we can simulate physiological and pathological scenarios on the same LA geometry, simply by acting on the 0D circulation model.  
\subsubsection{Lumped-parameter model}

The mathematical model that we use to derive the boundary conditions is a lumped-parameter model proposed in \cite{reg:primo}. It consists of a closed-loop 0D model, where geometrical reduction allows to represent the complete circulation in a synthetic way, considering only time-dependent variables such as pressures, fluxes, and volumes. The model describes the complete cardiovascular system, considering a subdivision into three main compartments: pulmonary, systemic, and cardiac circulation. The first two compartments are modelled through RLC circuit elements, while each cardiac chamber is represented by a capacitor with time-varying capacitance, called elastance. The heart valves are modeled via non-ideal diodes.
\begin{figure}[t]
	\centering
	{\includegraphics[width=0.65\textwidth]{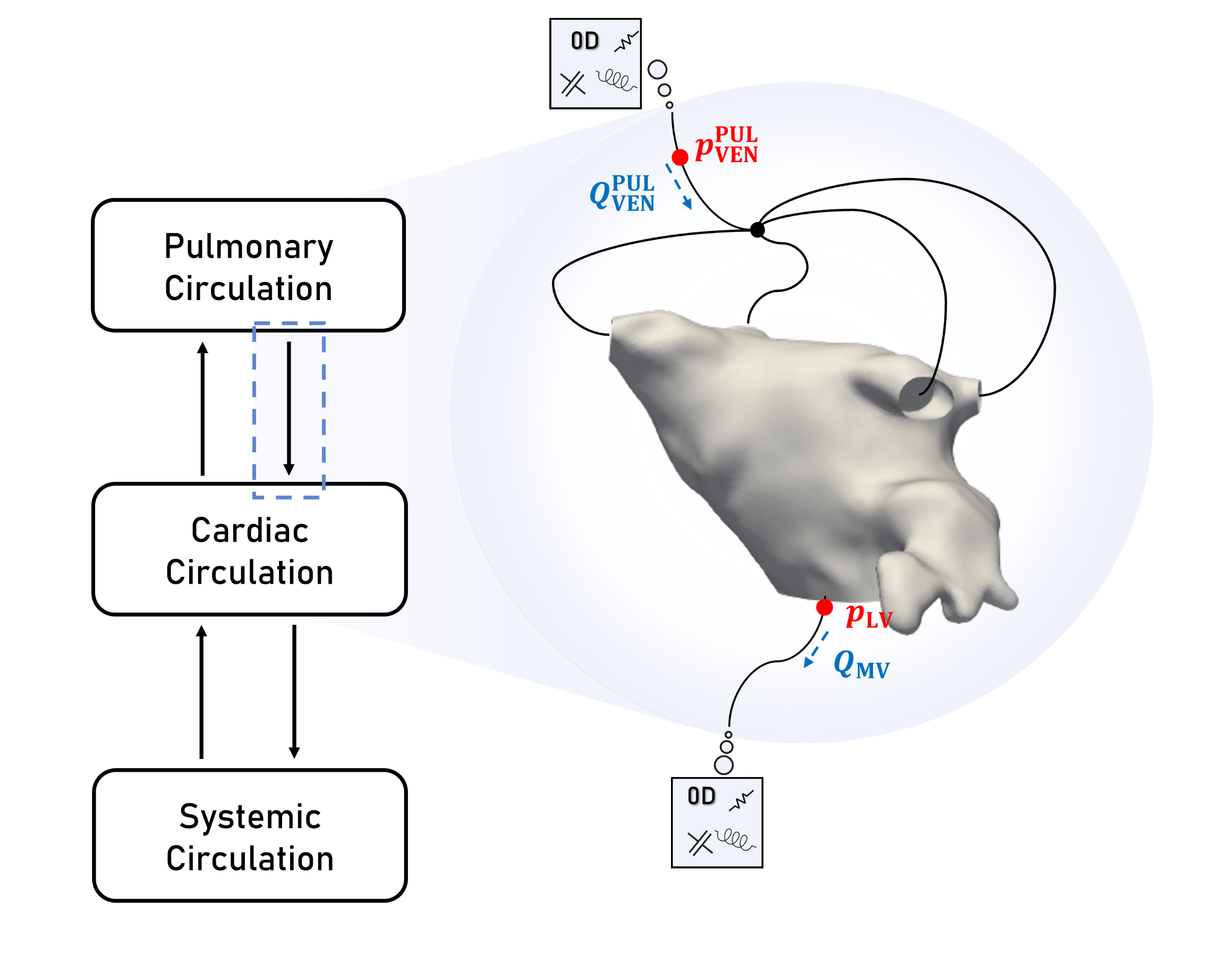}}
	\caption{\textcolor{black}{Relation between 0D variables and 3D domain in terms of boundary conditions that are prescribed to the 3D CFD problem. The coupling among the two models is enforced between the pulmonary and cardiac circulation. We highlight the left ventricular pressure $p_\mathrm{LV}$, mitral flux $Q_\mathrm{MV}$, pulmonary venous pressure $p_\mathrm{VEN}^\mathrm{PUL}$ and pulmonary venous flux $Q_\mathrm{VEN}^\mathrm{PUL}$.} }
	\label{fig:0Dimage}
\end{figure}
\par
We use the 0D model to calculate the fluid properties that serve as boundary conditions for the 3D fluid dynamics problem, as shown in Figure~\ref{fig:0Dimage}. The outlet pressure corresponds to the left ventricular pressure $p_\mathrm{OUT} = p_\mathrm{LV}$; the inlet pressure corresponds to the pulmonary venous pressure $p_\mathrm{IN} = p^\mathrm{PUL}_\mathrm{VEN}$. Concerning the Dirichlet inlet condition, we use the pulmonary veins flow rate $Q_\mathrm{IN} = Q^\mathrm{PUL}_\mathrm{VEN}$.
We first carry out a fully 0D simulation, then, once the 0D solution becomes periodic, we use pressures and flowrates transient to set the boundary conditions to the 3D CFD problem. Our approach can be regarded as a geometric multiscale problem, solved via a splitting algorithm \cite{quarteroni2016geometric}. 
\par
Moreover, the lumped-parameter model provides as output the volume of LA $V_\mathrm{LA}^\mathrm{0D}$, which is used to calibrate the displacement model, as we discuss in Section \ref{sec:disp}.
\subsubsection{Accounting for AF in the lumped-parameter circulation model}
\begin{figure}[t]
	\centering
	{\includegraphics[width=0.95\textwidth]{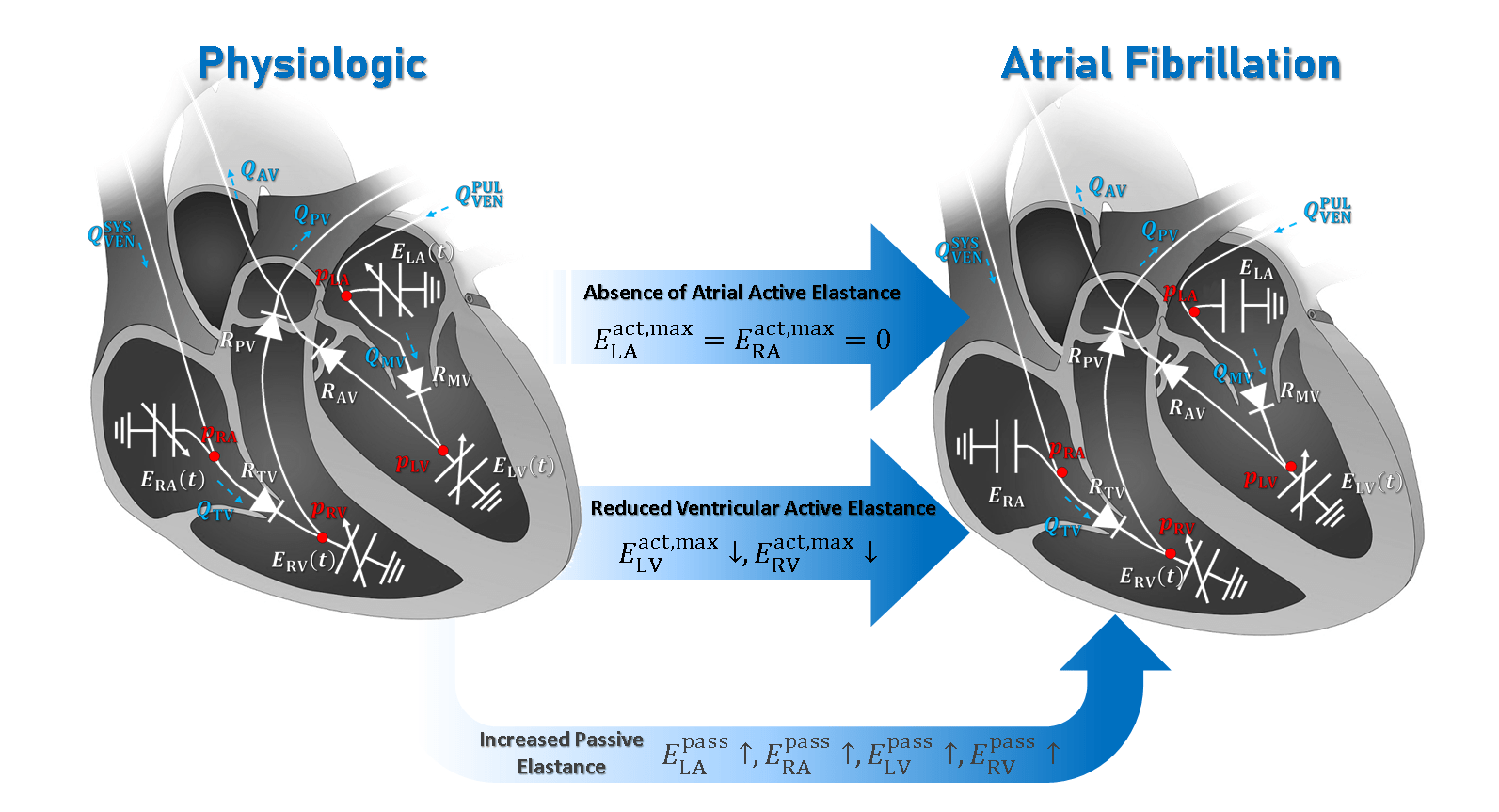}}
	\caption{\textcolor{black}{0D lumped parameter Model changes to simulate an AF pathology. On the left, the original circulation model, on the right the model with null atrial active elastances.The two corrections (increased passive elastance and reduced $E_\mathrm{RV}^\mathrm{act,max}$) are  necessary to fit the 3D patient-specific volumes. Variables names are coherent with model presented in \cite{reg:primo}.}}
	\label{fig:0DAF}
\end{figure}

The circulation model introduced in \cite{reg:primo} is tuned to model a healthy individual; for this reason, we calibrate the parameters to simulate the AF pathology correctly. To the best of our knowledge, the only existing work in which a lumped-parameter model is used to simulate the AF pathology is \cite{scarsoglio:AF}. The variations we use in this work are resumed in Figure~\ref{fig:0DAF}. We underline that the ones applied to passive and active elastances of the right ventricle are introduced by us to fit the correct 3D volumes of real patients, while all the others are also employed in \cite{scarsoglio:AF}.
\par

%\begin{equation}
%    E_j(t) = 
%    \begin{cases}
%        E_j^\mathrm{pass} + \dfrac{E_j^\mathrm{act,max}}{2}\left(1-\cos\left(\pi\dfrac{t}{T_\mathrm{d}}\right)\right) & 0 < t \leq T_\mathrm{d}, \\
%        E_j^\mathrm{pass} + \dfrac{E_j^\mathrm{act,max}}{2}\left(1+\cos\left(\pi\dfrac{t-T_\mathrm{d}}{T_\mathrm{HB}-T_\mathrm{d}}\right)\right) & T_\mathrm{d} < t \leq T_\mathrm{HB},
%    \end{cases}
%\end{equation}
To model the motion of the cardiac chambers, we use time-dependent elastances defined as \cite{reg:primo}. Each elastance can vary in the following prescribed range:
\begin{equation*}
	E_j(t) \in [E_j^{\mathrm{pass}},E_j^{\mathrm{pass}}+E_j^{\mathrm{act,max}}],\qquad j\in\{\mathrm{RA},\mathrm{LA},\mathrm{RV},\mathrm{LV}\},
\end{equation*}
being $E_j^\mathrm{pass}$, $E_j^\mathrm{act,max}$ the passive and active elastances, respectively .
\par
The effect of AF on the mechanical response of the cardiac tissue can be modelled by taking the active elastances equal to zero for both the atria, simulating hence the absence of the ``atrial kick'' \cite{iwasaki:AF,schotten:AF}, i.e. $
	E_\mathrm{LA}^\mathrm{act,max} = E_\mathrm{RA}^\mathrm{act,max} = 0$.
In AF, this choice implies a constant value for the elastances of the two chambers, namely:
\begin{equation*}
	E_\mathrm{LA}(t) = E_\mathrm{LA}^\mathrm{pass}, \qquad E_\mathrm{RA}(t) = E_\mathrm{RA}^\mathrm{pass} \qquad \forall t\in(0,T).
\end{equation*}
Under pathological transmission of the electric signal, we model the loss of ventricular contractility, reducing the active component of elastances in the ventricles.
\par
%However, we found that after these corrections we did not achieve meaningful results. In particular, we were not able to respect the constraint of the maximum volume of the atria given by the medical images that we are using. Indeed, 
To simulate the AF geometries, we increase the passive elastances of both atria and ventricles. This choice is fundamental to get the correct volumes and pulmonary venous pressure. Indeed, pulmonary hypertension has a connection with AF \cite{PVH,dodge:LV}, but without these corrections, the pressure values arising from the 0D model would become even higher than the pathological ones. After our calibration, the model also calculates lower values for left ventricular pressure $p_\mathrm{LV}$ than under physiological conditions, consistent with the pathological consequences of AF \cite{pappa:LV,dodge:LV}.
The passive elastance correction is smaller in atria affected by the remodeling, which caused a volume increase. In fact, this consequence of the pathology is typically detected by AF lumped-parameter model \cite{scarsoglio:AF}.
\subsubsection{Parametrization of the LA wall displacement}
\label{sec:disp}
Following \cite{zingaro:multiscale,zingaro:LA,dede:leftheart} we assume that the boundary datum $\boldsymbol{g}^\mathrm{ALE}$, introduced in Equation \ref{eq:harm}, can be expressed by means of the separation of variables as:
\begin{equation}
	\boldsymbol{g}^\mathrm{ALE}(\boldsymbol{x},t) = \boldsymbol{F}^\mathrm{ALE}(\boldsymbol{x})\;h^\mathrm{ALE}(t) \qquad \mathrm{on}\; \partial\Omega_t\times(0,T).
\end{equation}
In the following, we detail the construction of the two functions $h^\mathrm{ALE}$ and $\boldsymbol{F}^\mathrm{ALE}$.
%\subsubsection{Time-dependent function component}
Let $V_\mathrm{LA}(t)$ be the LA volume; then, by using the Reynolds transport theorem (RTT) \cite{kundu:CFD}:
\begin{equation}
	\dfrac{\,\mathrm{d}V_\mathrm{LA}}{\mathrm{d}t} = \dfrac{\mathrm{d}}{\mathrm{d}t}\int_{\Omega_t} \mathrm{d}\boldsymbol{x} \overset{\mathrm{(RRT)}}{=} \int_{\partial\Omega_t} \boldsymbol{g}^\mathrm{ALE}\cdot \boldsymbol{n} \mathrm{d}\sigma = 
	 h^\mathrm{ALE}\int_{\partial\Omega_t} \boldsymbol{F}^\mathrm{ALE}\cdot \boldsymbol{n} \mathrm{d}\sigma,
\end{equation}
we obtain the following definition of the time-dependent function:
\begin{equation}
	h^\mathrm{ALE}(t) = \Bigg(\int_{\partial\Omega_t} \boldsymbol{F}^\mathrm{ALE}\cdot \boldsymbol{n}(t) d\sigma\Bigg)^{-1}\dfrac{\,\mathrm{d}V_\mathrm{LA}(t)}{\mathrm{d}t}.
\end{equation}
Specifically, we set the LA volume to be equal to that computed via the 0D circulation model ($V_\mathrm{LA}(t) = V_\mathrm{LA}^\mathrm{0D}(t)$) for each time $t\in(0,T)$.
\textcolor{black}{This choice, together with the enforcement of boundary conditions provided from the circulation model, ensures that the compatibility of NS boundary conditions \cite{quartapelle:CFD} is automatically satisfied by the mass conservation property of the 0D model \cite{reg:primo}.}
\subsubsection{Space-dependent function and LAA correction}
The space-dependent component of the boundary function $\boldsymbol{F}^\mathrm{ALE}$ is constructed by considering two different components on the corresponding reference domain:
\begin{equation}
	\widehat{\boldsymbol{F}}^\mathrm{ALE}(\widehat{\boldsymbol{x}}) = \widehat{\boldsymbol{f}}^\mathrm{ALE}(\widehat{\boldsymbol{x}}) + \widehat{\boldsymbol{f}}_\mathrm{LAA}^\mathrm{ALE}(\widehat{\boldsymbol{x}}).
	\label{eq:separ}
\end{equation}
In Equation \ref{eq:separ}, we separate the motion of the whole chamber, modelled by $\widehat{\boldsymbol{f}}^\mathrm{ALE}$, from the one of the LAA, which is separately described by $\widehat{\boldsymbol{f}}_\mathrm{LAA}^\mathrm{ALE}$. A correct calibration of these two functions generates a displacement that can be adapted to the LAA morphology. Moreover, by introducing this distinction, we have control on the LAAEF, which can be set to be coherent with the clinical measurements found in literature concerning the pathologic situation of the patient, providing hence a better estimate of the blood motion inside the most dangerous region in terms of thrombous formation \cite{LAA:al-saady,LAA:ernst,LAA:tan}.
\par
\begin{figure}[t]
	\centering
	\includegraphics[width=0.95\textwidth]{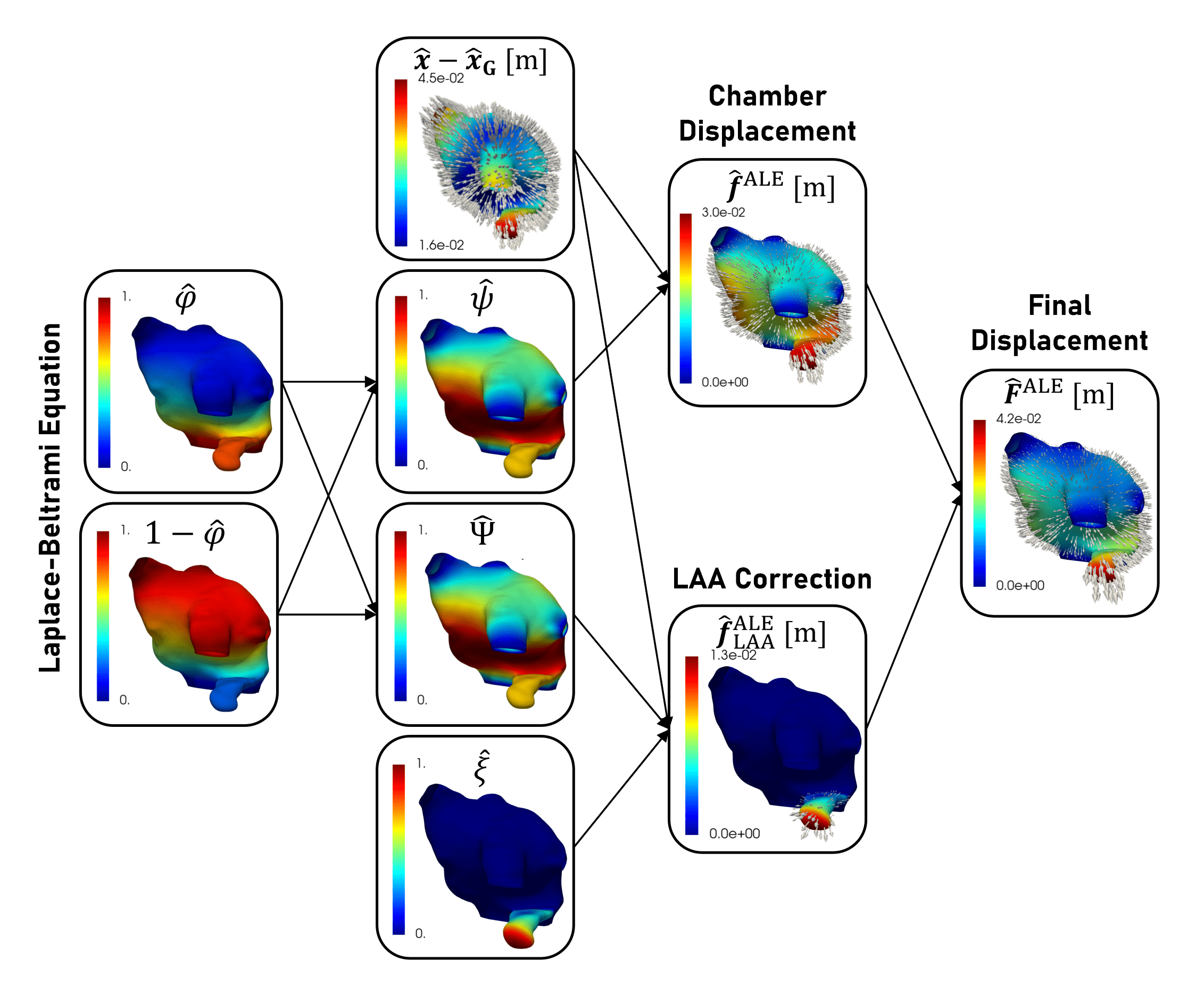}
	\caption{\textcolor{black}{Complete procedure to derive the spatial component of the boundary displacement  $\boldsymbol{F}^\mathrm{ALE}$. We start from the Laplace-Beltrami solution $\widehat{\varphi}$, the LAA smoothing function $\widehat{\xi}$ and the distance from centre of mass $\widehat{\boldsymbol{x}}-\widehat{\boldsymbol{x}}_\mathrm{G}$. The image shows also the partial results needed: $\widehat{\psi}$, $\widehat{\Psi}$, $\boldsymbol{f}^\mathrm{ALE}$,$\boldsymbol{f}_\mathrm{LAA}^\mathrm{ALE}$}}
	\label{fig:procedurefale}.
\end{figure}

We define the global component $\widehat{\boldsymbol{f}}^\mathrm{ALE}$ as:
\begin{equation}
	\widehat{\boldsymbol{f}}^\mathrm{ALE}(\widehat{\boldsymbol{x}}) = \widehat{\psi}(\widehat{\boldsymbol{x}})\;\left(\widehat{\boldsymbol{x}}-\widehat{\boldsymbol{x}}_\mathrm{G}\right),
\end{equation}
where the second term is a vector field directed to the centre of mass of the atrium $\widehat{\boldsymbol{x}}_\mathrm{G}$. We compute the function $\widehat \psi$ as a normalised product:
\begin{equation}
	\widehat{\psi}(\widehat{\boldsymbol{x}}) = \dfrac{\widehat{\varphi}(\widehat{\boldsymbol{x}})\;\left(1-\widehat{\varphi}(\widehat{\boldsymbol{x}})\right)}{\underset{\widehat{\boldsymbol{x}}\in\partial\widehat{\Omega}}{\max}\left\{\widehat{\varphi}(\widehat{\boldsymbol{x}})\;\left(1-\widehat{\varphi}(\widehat{\boldsymbol{x}})\right)\right\}},
\end{equation}
being $\widehat{\varphi}$ solution of the following Laplace-Beltrami problem \cite{Bonito:2020}:
\begin{equation}
	\begin{cases}
		-\Delta_\Gamma \widehat{\varphi} = 0 & \mathrm{in}\;\partial\widehat{\Omega}, \\
		\widehat{\varphi} = 0 		& \mathrm{on}\;\bigcup_{j=1}^{4}\widehat{\Gamma}^{\mathrm{PV}_j}, \\
		\widehat{\varphi} = 1 		& \mathrm{on}\;\widehat{\Gamma}^\mathrm{MV}.
	\end{cases}
\end{equation}
By defining $\widehat{\psi}: \partial\widehat{\Omega} \to [0, 1]$ as described, we get a smooth function which is zero on the inlet and outlet sections of our computational domain, and non-null in the main chamber, as displayed in Figure~\ref{fig:procedurefale}. 
\par
Analogously, we construct the function $\widehat{\boldsymbol{f}}_\mathrm{LAA}^\mathrm{ALE}$ as follows:
\begin{equation}
	\widehat{\boldsymbol{f}}_\mathrm{LAA}^\mathrm{ALE}(\widehat{\boldsymbol{x}}) = \widehat{\Psi}(\widehat{\boldsymbol{x}})\;\widehat{\xi}(\widehat{\boldsymbol{x}})\;(\widehat{\boldsymbol{x}}-\widehat{\boldsymbol{x}}_\mathrm{G}^\mathrm{LAA}),
\end{equation}
where $\widehat{\boldsymbol{x}}_\mathrm{G}^\mathrm{LAA}$ is the center of mass of the LAA and we define it as $\widehat{\Psi}(\widehat{\boldsymbol{x}}) = k\;\widehat{\psi}(\widehat{\boldsymbol{x}})$. The use of a multiplicative constant $k$ allows us to vary the magnitude of the LAA contraction, according to the LAAEF, which characterizes the pathological situation of the patient. By defining $\widehat{\boldsymbol{f}}_\mathrm{LAA}^\mathrm{ALE}$ as explained, we get a displacement which is directed towards the center of mass of the LAA; moreover, it is fundamental the use of a function $\widehat{\xi}$ to localize the support only at the LAA surface, indeed the changes need to be located only in this region\footnote{We remark that by using an identity function to localize the support may cause discontinuities which can lead to a ``break'' of the surface, for this reason we apply a mollifier to avoid this problem.}.
\subsection{Setup of numerical simulations}
\label{sec:NSS}
We carry out numerical simulations on four ideal patients. In particular, we considered a LA geometry related to a subject with physiological conditions, assuming first sinus rhythm and then AF, and two geometries of patients really affected by AF. The imaging method used to reconstruct medical images is the diffusion tensor magnetic resonance imaging, which allows better resolution of the thin atrial wall \cite{pash:geom}. The hearts were of donors from National Disease Research Interchange (Philadelphia, PA). Geometric and clinical information for the three patients are resumed in Table~\ref{tab:geomdata}. The volumes of patients AF2 and AF3 are larger than those normally detected under physiological conditions. These values suggest an atrial remodeling caused by the AF pathology; hinting at the possibility of a persistent AF.

\begin{table}[t]
\centering
\begin{tabular}{|l|c|c|c|}
\hline
\textbf{Geometry}                     
& \textbf{P1}     & \textbf{P2}     & \textbf{P3}  \\ \hline
\textbf{Age}                 
& $55$ Years      & $86$ Years     & $94$ Years    \\ \hline
\textbf{Gender}              
& Male          & Male         & Female       \\ \hline
\textbf{Pathology}         
& None        & AF          & AF      \\ \hline
\textbf{LA Max Volume} [$\mathrm{cm}^3$]           
& $58.06$    & $85.55$    & $69.39$       \\ 
\hline
\textbf{RA Max Volume} [$\mathrm{cm}^3$]           
& $69.39$   & $55.76$     & $36.33$        \\ 
\hline
\textbf{ID in repository \cite{roney:geom}}         
& $7$   & $6$     & $5$    \\ 
\hline
\end{tabular}
\smallskip
\caption{Information of the geometries detected from images.}
\label{tab:geomdata}
\end{table}

\subsubsection{Meshes}
\begin{figure}[t]
	\centering
	{\includegraphics[width=0.95\textwidth]{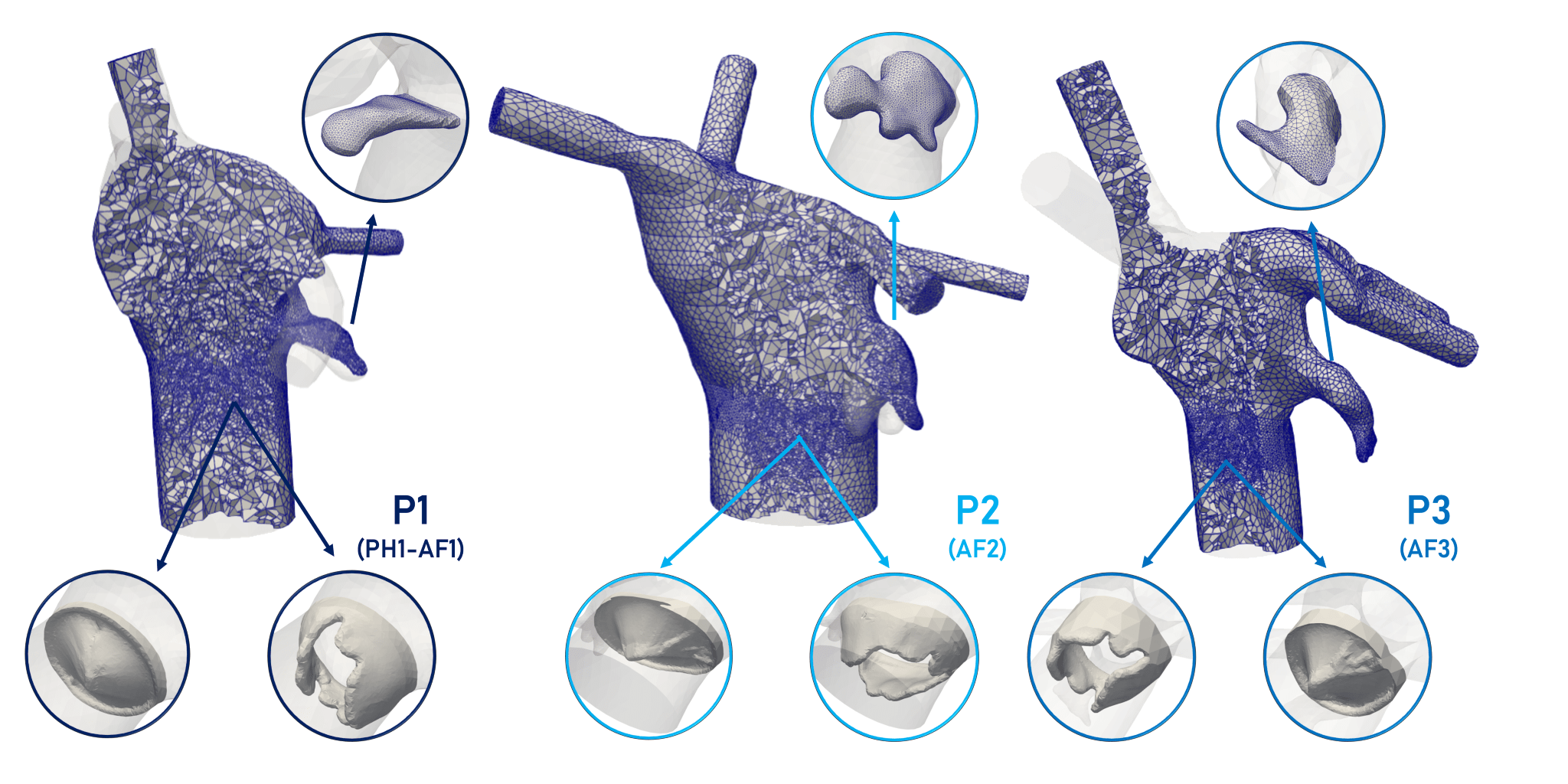}}
	\caption{\textcolor{black}Hexaedral meshes used to simulate the patients haemodynamics, with the detail of: LAA morphology, the MV in open and closed configurations. Meshes are refined in the LAA region, to correctly capture the geometrical features, and near the MV, in order to represent its leaflet  by means of the RIIS method.}
	\label{fig:meshes}
\end{figure}

As we can see in Figure~\ref{fig:meshes}, we build a hexahedral mesh with a heterogeneous size $h_K$. We perform a refinement in the LAA to capture the geometrical features of this region and near the MV to correctly represent the valve using the RIIS method. We use a value of $\varepsilon = 1.3\;\mathrm{mm}$ due to medical estimates of $2.6\; \mathrm{mm}$ of the thickness of the MV leaflet \cite{gray:anatomy,MV:MI} ($\varepsilon$ represents the half-thickness of the valve leaflets). Concerning opening $\Delta\tau_\mathrm{O}$ and closure $\Delta\tau_\mathrm{C}$ duration of the MV, we consider literature estimates, and we set $\Delta\tau_\mathrm{O} = 20\;\mathrm{ms}$ \cite{MV:open} and $\Delta\tau_\mathrm{C} = 60\;\mathrm{ms}$ \cite{MV:close}. We consider the opening of the valve when the pressure of a control volume inside the LA is higher than the one computed inside a volume downwind the valve. The valve closure starts at a fixed time, imposed when $\dot{V}_\mathrm{LV}$ of the 0D model simulation becomes negative, i.e. when a condition of reversed flow is detected on the outlet section \cite{Fer:RIS}.
\par
The MV geometries were not available in the repository \cite{niederer:geom, roney:geom}. For this reason, we adapt the valve geometry provided by Zygote \cite{zygote:report} to the orifice of the patient-specific LA. Moreover, the leaflet displacement we prescribe is the same that has been designed in \cite{zingaro:multiscale}. 
\par
Concerning the pulmonary veins, we add some rigid tubes to the domain to obtain at the LA inlet a fully developed flow and to reduce the influence of the parabolic profile choice.
\par
Using the VMTK library \cite{VMTK:antiga}, we first generate a tetrahedral mesh, then we use mesh tethex
\cite{tethex} to obtain a hexaedral mesh in which each tetrahedron is split into four hexahedra preserving the
aspect ratio of the original element. Information about the constructed meshes is reported in Table~\ref{tab:meshdata}. A more detailed description of the preprocessing of tools we used to generate the meshes can be found in \cite{Corti:Tesi}. 
\begin{table}[t]
\centering

\begin{tabular}{|c|r|r|r|}
\hline
\textbf{Mesh}   
& \multicolumn{1}{c|}{\textbf{P1}}
& \multicolumn{1}{c|}{\textbf{P2}}
& \multicolumn{1}{c|}{\textbf{P3}}  \\
\textbf{(Patient)}   
& \multicolumn{1}{c|}{\textbf{(PH1-AF1)}}
& \multicolumn{1}{c|}{\textbf{(AF2)}}
& \multicolumn{1}{c|}{\textbf{(AF3)}}  \\ \hline
\textbf{Number of Cells}                 
& $120\,860$    & $182\,540$    & $110\,472$ \\ \hline
\textbf{Velocity $\mathrm{DOFs}$}              
& $411\,873$    & $614\,631$    & $378\,879$ \\ 
\textbf{Pressure $\mathrm{DOFs}$}              
& $137\,291$    & $204\,877$    & $126\,293$ \\
\textbf{Total $\mathrm{DOFs}$}              
& $549\,164$    & $819\,508$    & $505\,172$
\\ 
\hline
\hline
\textbf{$h_K^\mathrm{max}\;[\mathrm{mm}]$}              
& $5.95$    
& $5.44$     
& $6.15$    \\ 
\textbf{$h_K^\mathrm{min}\;[\mathrm{mm}]$}              
& $0.43$
& $0.40$     
& $0.38$    \\ 
\textbf{$h_K^\mathrm{avg}\;[\mathrm{mm}]$}              
& $1.76$
& $1.69$     
& $1.82$    \\ \hline
\end{tabular}
\smallskip
\caption{\textcolor{black}{Details of the hexaedral meshes generated to carry out CFD simulations. The DOFs are referred to linear finite elements for velocity and pressure. }}
\label{tab:meshdata}
\end{table}
\subsubsection{CFD simulations}
All the simulations, based on the mathematical models we have shown, have been executed in $\texttt{life}^\texttt{x}$ \cite{lifex, lifex_core}, a high-performance \texttt{C++} FE library developed within the iHEART
project\footnote{iHEART - An Integrated Heart model for the simulation of the cardiac function, European Research Council (ERC) grant agreement No 740132, P.I. Prof. A. Quarteroni, 2017-2022}, mainly focused on cardiac simulations and based on \texttt{deal.II} finite element core \cite{dealii}. Numerical simulations are carried out on the cluster of the Department of Mathematics, Politecnico di Milano.  Specifically, the simulations PH1, AF1, and AF2 were run in the Gigat queue (\textit{6 nodes, 12 Intel Xeon E5-2640v4 @ 2.40GHz, 120 cores, 384GB RAM, O.S. Centos 6.7}) using 2 nodes with 20 cores each, and AF3 was run in the Gigatlong cluster (\textit{5 nodes, 10 Intel Xeon Gold 6238 @ 2.10GHz, 280 cores, 2.5TB RAM}) using 1 node with 56 cores.
\par
Blood density and dynamic viscosity are set equal to $\rho = 1.06\cdot 10^{3}\; \mathrm{kg/m^3}$ and $\mu = 3.5\cdot 10^{-3}\; \mathrm{kg/(m\cdot s)}$, respectively. We simulate six heartbeats, of period $T_\mathrm{HB}=1\,\mathrm{s}$, starting from a null initial condition. However, to filter out the unphysical consequences of this choice, we discard the first two heartbeats and we consider the phase-averaged velocity, defined over $N_\mathrm{HB}$ heartbeats\footnote{We simulate six heartbeats sincewe are interested in the phase-averaged fluid properties. In particular, we choose $N_\mathrm{HB} = 4$ to limit the computational cost.}, as:
\begin{equation}
	\langle\boldsymbol{u}(\boldsymbol{x},t)\rangle = \dfrac{1}{N_\mathrm{HB}}\sum_{n = 1}^{N_\mathrm{HB}} \boldsymbol{u}(\boldsymbol{x},t+(n-1)T_\mathrm{HB}).
\end{equation}
\subsubsection{Lagrangian simulations}
\label{sec:lag-explanation}
The use of Lagrangian simulations to detect indicators can provide additional information on the haemodynamics \cite{chnafa:LES, Rossini2016ACM}. We perform numerical simulations of the movement of red blood cells inside the LA, using as velocity field the result of the NS simulations. In addition, we consider the introduction of $N_H$ parcels\footnote{A parcel $j$ is a macro-particle associated to a number $\omega_j$ of real particles, in our case red blood cells. This numerical approximation is typical of the Discrete Parcel Method (DPM) \cite{DPM}. A complete explanation of the concept can be found in the supplementary material} in each heartbeat. In particular, we consider the injection of a volume of blood $V_\mathrm{Inj} = \int_0^{T_\mathrm{HB}} Q_\mathrm{VEN}^\mathrm{PUL}(t) \mathrm{d}t$.
\par
We use the injected volume to estimate the number $N_P$ of cells entering the atrium during a single heartbeat, also considering that the number of red blood cells in a cubic millimetre is approximately $5$ millions \cite{dean:RBC}. Our simulation is designed to obtain a constant approximate weight $\omega = N_P/N_H  \simeq 1409$, common to all cases, where $N_P$ is the number of simulated red blood cells.
\par
\begin{figure}[t]
	\centering
	{\includegraphics[width=0.8\textwidth]{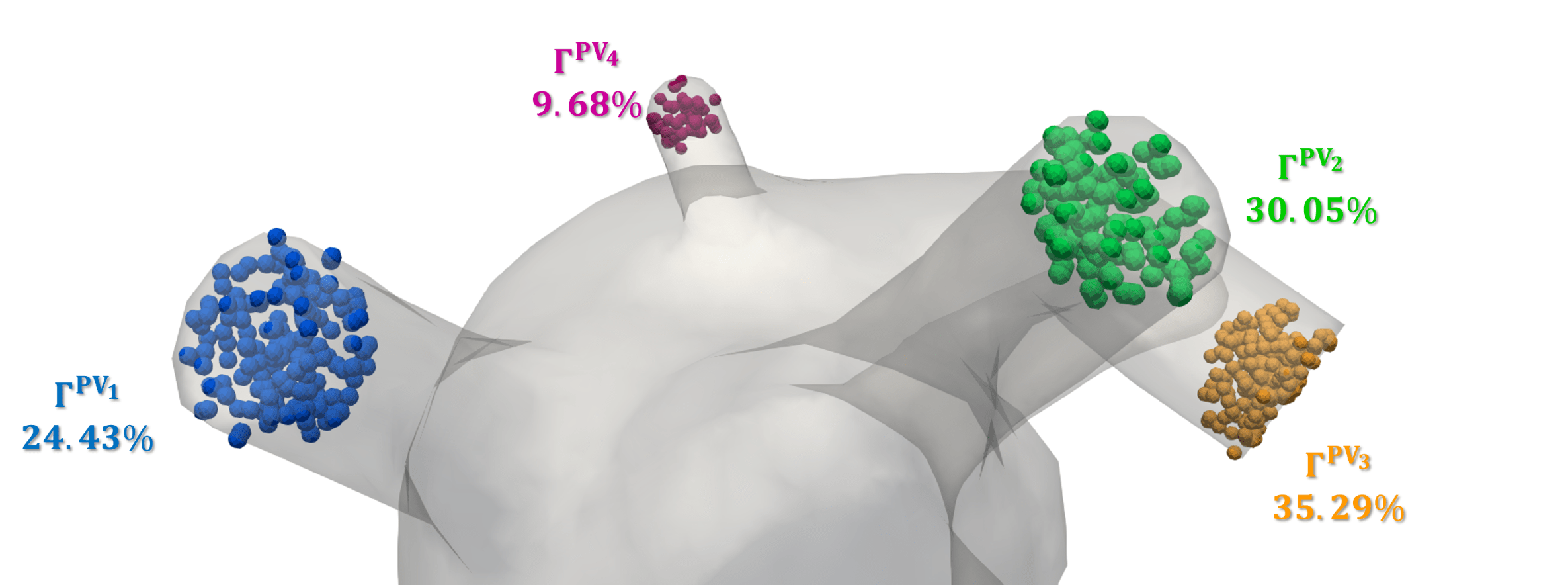}}
	\caption{Position of parcels injection in cases PH1 and AF1 associated to the inlet sections $\Gamma^{PV}_j$ and to the flow rate $\zeta_j$ in Equation \ref{eq:repartition}.}
	\label{fig:inj}
\end{figure}

The number of parcels injected at any timestep can be computed as follows:
%\begin{equation}
$
    n_H(t) = \dfrac{Q_\mathrm{VEN}^\mathrm{PUL}(t)}{V_\mathrm{Inj}}\,\Delta\tau\, N_\mathrm{H},
$
%\end{equation}
where $\Delta\tau$ is the time step we choose for the Lagrangian simulation. We split the injection among the four pulmonary veins according to the flow repartition factor introduced in Equation \ref{eq:repartition}; the parcels are distributed in a hemisphere centered at the beginning of the pulmonary vein extension and with the same radius of the cylinder. The initial position of the parcels is randomly chosen within the hemisphere, and an example can be seen in Figure~\ref{fig:inj}.
\par 
More information on the equation of motion of the parcels and the construction of Eulerian fields from the Lagrangian perspective can be found in the supplementary material and in \cite{Corti:Tesi}. This type of modelling is a first attempt to detect the indicators to estimate thrombotic risk. As we are aware of the limitations of this indicator, we plan to further investigate it in the future. For example, we can improve the equations of motion (i.e. introducing wall attachment properties as in \cite{planas:1}).

\section{Results}
\label{sec:results}
\textcolor{black}{In this Section, we show the results of numerical simulations. Specifically, we present the results of the circulation model in Section~\ref{sec:results-0d} and the 3D Eulerian and Lagrangian indicators in Sections~\ref{sec:eul} and \ref{sec:lag}, respectively. Section~\ref{sec:agestasis} is devoted to the introduction of the new hemodynamic indicator.   }
\subsection{Results of lumped-parameter model}
\label{sec:results-0d}

The parameter values used in the simulation of the lumped-parameter model have to be calibrated starting from values present in the literature \cite{heldt:0D,reg:primo,liang:0D,scarsoglio:AF}. However, due to the requirement to fit the information from the medical images and the pathological situation, they need to be separately tuned for each patient. We report the parameters that we keep common to all cases in Table~\ref{tab:common0Dparam}. Instead, in Table~\ref{tab:0Dela}, we store the chamber elastances in the four simulated cases. We choose the elastances starting from literature values in \cite{reg:primo}, and they are manually calibrated to fit the volume of the atrial chambers and to reach reasonable ejection fractions.
\par
In Table~\ref{tab:patientdata}, we list the Left Atrial Ejection Fraction (LAEF) and LAAEF, calculated after our procedure and defined as:
\begin{equation}
    \mathrm{LAEF} = \dfrac{V_\mathrm{LA}^\mathrm{max}-V_\mathrm{LA}^\mathrm{min}}{V_\mathrm{LA}^\mathrm{max}}, \quad \mathrm{and} \quad
    \mathrm{LAAEF} = \dfrac{V_\mathrm{LAA}^\mathrm{max}-V_\mathrm{LAA}^\mathrm{min}}{V_\mathrm{LAA}^\mathrm{max}},
\end{equation}
respectively. We remark that the maximum volumes are directly retrieved by medical images; whereas the minimum ones are determined after the application of the displacement. The in-silico values we compute are in accordance with the clinical measurements, for both LAEF \cite{LAEF:appleton,LAEF:kaufmann}, and LAAEF \cite{LAAEF:gan}, making the whole displacement procedure significant and reliable. 
\begin{table}[t]
\centering
\begin{tabular}{|c|c|c|c|c|c|}
\hline
\textbf{Patient}     
& \multirow{2}{*}{\textbf{Pathology}}    
& \multirow{2}{*}{\textbf{Indicator}}
& \textbf{Simulated}    
& \textbf{Clinical} 
& \multirow{2}{*}{\textbf{Reference}} \\
\textbf{(Geometry)}   
& 
& 
& \textbf{value}
& \textbf{measurement} 
& \\
\hline
\begin{tabular}[c]{@{}c@{}}\textbf{PH1}  \\ (\textbf{P1})\end{tabular}    & None   
& \begin{tabular}[c]{@{}c@{} }LAEF (\%)  \\ LAAEF (\%)\end{tabular}    
& \begin{tabular}[c]{@{}c@{}} $50.49$\\ $77.81$ \end{tabular}  
& \begin{tabular}[c]{@{}c@{}} $45.1\div64.1$ \\ $62.5\div86.5$ \end{tabular}  
& \begin{tabular}[c]{@{}c@{}} \cite{LAEF:appleton} \\ \cite{LAAEF:gan} \end{tabular}  \\ \hline
\begin{tabular}[c]{@{}c@{}}\textbf{AF1}  \\ (\textbf{P1})\end{tabular}      & \begin{tabular}[c]{@{}c@{}}Paroxysmal  \\ AF\end{tabular}        
& \begin{tabular}[c]{@{}c@{}} LAEF (\%)  \\ LAAEF (\%)\end{tabular}    
& \begin{tabular}[c]{@{}c@{}} $30.72$ \\ $62.72$  \end{tabular}
& \begin{tabular}[c]{@{}c@{}} $10.0\div50.0$ \\  $50.0\div69.0$ \end{tabular}  
& \begin{tabular}[c]{@{}c@{}} \cite{LAEF:kaufmann} \\ \cite{LAAEF:gan} \end{tabular} \\ \hline
\begin{tabular}[c]{@{}c@{}}\textbf{AF2}  \\ (\textbf{P2})\end{tabular}      & \begin{tabular}[c]{@{}c@{}}Persistent  \\ AF\end{tabular}         
& \begin{tabular}[c]{@{}c@{}} LAEF (\%) \\ LAAEF (\%)\end{tabular}   
& \begin{tabular}[c]{@{}c@{}} $18.48$ \\ $38.94$  \end{tabular}
& \begin{tabular}[c]{@{}c@{}} $6.0\div26.0$ \\ $32.4\div63.2$ \end{tabular}  
& \begin{tabular}[c]{@{}c@{}} \cite{LAEF:kaufmann} \\ \cite{LAAEF:gan} \end{tabular} \\ \hline
\begin{tabular}[c]{@{}c@{}}\textbf{AF3}  \\ (\textbf{P3})\end{tabular}     & \begin{tabular}[c]{@{}c@{}}Persistent  \\ AF\end{tabular}       
& \begin{tabular}[c]{@{}c@{}} LAEF (\%)  \\ LAAEF (\%)\end{tabular}    
& \begin{tabular}[c]{@{}c@{}} $21.17$ \\ $53.45$ \end{tabular}
& \begin{tabular}[c]{@{}c@{}} $6.0\div26.0$ \\ $32.4\div63.2$ \end{tabular}  
& \begin{tabular}[c]{@{}c@{}} \cite{LAEF:kaufmann} \\ \cite{LAAEF:gan} \end{tabular} \\ \hline
\end{tabular}
\smallskip
\caption{\textcolor{black}{Patient information from images and modelling assumptions with simulated values of atrial and auricle ejection fractions, clinical measurement are intended in a range (min $\div$ max).}}
\label{tab:patientdata}
\end{table}
\begin{figure}[t]
	\centering
	{\includegraphics[width=0.95\textwidth]{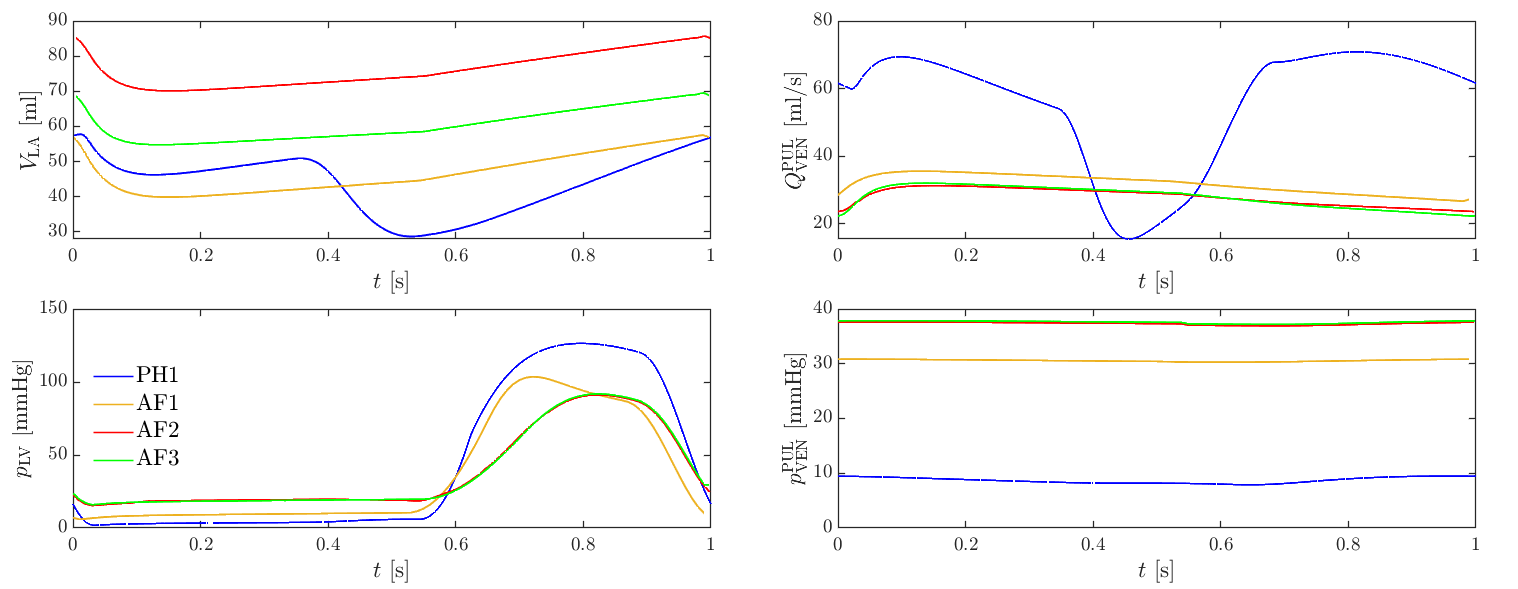}}
	\caption{\textcolor{black}{Results of the lumped-parameter model  simulations for the four patients: left atrial volume $V_\mathrm{LA}$ (top, left), pulmonary venous flux $Q_\mathrm{VEN}^\mathrm{PUL}$ (top, right), left ventricular pressure $p_\mathrm{LV}$ (down, left) and pulmonary venous pressure $p_\mathrm{VEN}^\mathrm{PUL}$ (down, right). These transients are then used as boundary data for the 3D CFD simulations.}}
	\label{fig:LPMBCs}
\end{figure}
\par
We report the volume of LA, the pulmonary venous flow rate and pressure, and the ventricular pressure in Figure~\ref{fig:LPMBCs}, for the four simulated cases. We recall that these functions are taken as output from the 0D model and used as boundary data for our 3D CFD simulation. 

\subsection{CFD Eulerian indicators}
\label{sec:eul}

\begin{figure}[t]
	\centering
	{\includegraphics[width=0.95\textwidth]{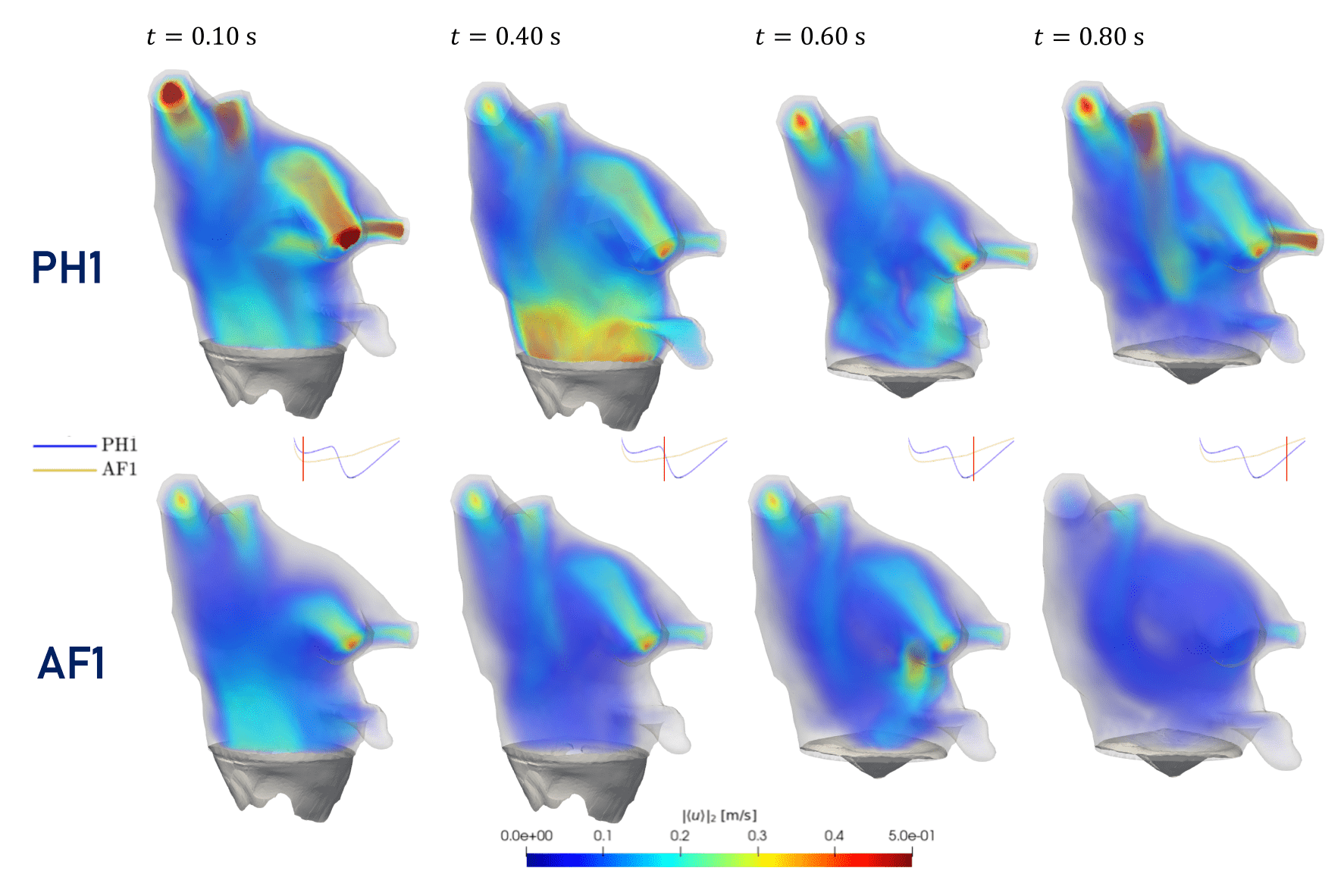}}
	\caption{\textcolor{black}{Volume rendering of phase-averaged velocity magnitude $|\langle\boldsymbol{u}\rangle|_2$ at different time instants in physiologic and AF conditions considering the same atrial geometry. The time instants are reported on the volume transients to distinguish diastolic and systolic phases.}}
	\label{fig:velfield}
\end{figure}

We report the velocity magnitude in Figure~\ref{fig:velfield} in a single heartbeat. The velocity magnitude is larger in physiologic conditions than in AF ones during the whole heartbeat. This behaviour is evident during the E-wave ($t\simeq0.10\,\mathrm{s}$), whereas the A-wave is present only in the patient PH1 ($t=0.40\,\mathrm{s}$). When the MV is closed, we can notice the difference in contraction of the two atrial chambers, due to the contractile reduction in fibrillation ($t\simeq0.60\,\mathrm{s}$). 

\begin{table}[t]
\centering
\begin{tabular}{|c|c|c|c|c|}
\hline
\multirow{2}{*}{\textbf{Patient}}     & \multirow{2}{*}{\textbf{Indicator}}
& \textbf{\textit{In silico}}    & \textbf{Clinical} & \multirow{2}{*}{\textbf{Reference}} \\ 
& & \textbf{result} & \textbf{measurement} & \\ \hline 
\textbf{PH1}     
&   \begin{tabular}[l]{@{}l@{}l@{}l@{}l@{}l@{}l@{}} 
        Peak velocity veins [m/s]  \\ 
        Peak velocity [m/s] \\
        Mean velocity [m/s] \\
        E-peak velocity [m/s] \\ 
        A-peak velocity [m/s] \\ 
        E/A ratio [-] \\
        Peak LAA emptying velocity [m/s]
    \end{tabular}    
&   \begin{tabular}[c]{@{}c@{}c@{}c@{}c@{}c@{}c@{}} 
        $0.474$ \\ 
        $0.775$ \\
        $0.149$ \\
        $0.918$ \\ 
        $0.544$ \\ 
        $1.680$ \\
        $0.428$
    \end{tabular}  
&   \begin{tabular}[c]{@{}c@{}c@{}c@{}c@{}c@{}c@{}} 
        $0.370\div0.600$ \\ 
        $0.560\div1.060$ \\ 
        $0.130\div0.170$ \\
        $0.740\div1.040$ \\ 
        $0.520\div1.040$ \\ 
        $0.800\div1.870$ \\ 
        $0.279\div0.432$
    \end{tabular}  
&   \begin{tabular}[c]{@{}c@{}c@{}c@{}c@{}c@{}c@{}} 
        \cite{4DMRI:garcia}  \\ 
        \cite{4DMRI:garcia}  \\ 
        \cite{4DMRI:markl}   \\
        \cite{E/A:phys}      \\ 
        \cite{E/A:phys}      \\ 
        \cite{E/A,E/A:phys}  \\
        \cite{4DMRI:marklLAA}
    \end{tabular}
\\ 
\hline
\textbf{AF1}      
&   \begin{tabular}[l]{@{}l@{}l@{}l@{}l@{}} 
        Peak velocity veins [m/s]  \\ 
        Peak velocity [m/s] \\
        Mean velocity [m/s] \\
        E-peak velocity [m/s] \\
        Peak LAA emptying velocity [m/s]
    \end{tabular}  
&   \begin{tabular}[c]{@{}c@{}c@{}c@{}c@{}} 
        $0.276$ \\ 
        $0.480$ \\ 
        $0.096$ \\
        $0.612$ \\
        $0.413$
    \end{tabular}
&   \begin{tabular}[c]{@{}c@{}c@{}c@{}c@{}} 
        $0.250\div0.540$ \\  
        $0.470\div0.970$ \\ 
        $0.080\div0.140$ \\
        $0.443\div0.771$ \\
        $0.208\div0.428$ 
    \end{tabular}  
&   \begin{tabular}[c]{@{}c@{}c@{}c@{}c@{}} 
        \cite{4DMRI:garcia} \\ 
        \cite{4DMRI:garcia} \\ 
        \cite{4DMRI:markl} \\
        \cite{E/A:korea} \\
        \cite{4DMRI:marklLAA}
    \end{tabular}
\\ 
\hline
\textbf{AF2}    
& \begin{tabular}[l]{@{}l@{}} 
        Peak velocity veins [m/s]  \\ 
        Peak velocity [m/s] \\
        Mean velocity [m/s] \\
        E-peak velocity [m/s] \\
        Peak LAA emptying velocity [m/s]
    \end{tabular}   
&  \begin{tabular}[c]{@{}c@{}} 
        $0.288$ \\
        $0.177$ \\
        $0.058$ \\
        $0.664$ \\
        $0.366$
    \end{tabular}  
& \begin{tabular}[c]{@{}c@{}} 
        - \\
        $0.150\div0.290$ \\
        $0.056\div0.160$ \\
        $0.516\div0.812$ \\
        $0.160\div0.390$
    \end{tabular}     
& \begin{tabular}[c]{@{}c@{}} 
        - \\
        \cite{4DMRI:marklLAA} \\
        \cite{4DMRI:marklLAA} \\
        \cite{E/A:korea} \\
        \cite{4DMRI:marklLAA}
    \end{tabular}   \\
\hline
\textbf{AF3}    
& \begin{tabular}[l]{@{}l@{}} 
        Peak velocity veins [m/s]  \\ 
        Peak velocity [m/s] \\
        Mean velocity [m/s] \\
        E-peak velocity [m/s] \\
        Peak LAA emptying velocity [m/s]
    \end{tabular}   
&  \begin{tabular}[c]{@{}c@{}} 
        $0.149$ \\
        $0.217$ \\
        $0.065$ \\
        $0.591$ \\
        $0.347$
    \end{tabular}  
& \begin{tabular}[c]{@{}c@{}}         
        - \\
        $0.150\div0.290$ \\
        $0.056\div0.160$ \\
        $0.516\div0.812$ \\
        $0.160\div0.390$
    \end{tabular}     
& \begin{tabular}[c]{@{}c@{}} 
        - \\
        \cite{4DMRI:marklLAA} \\
        \cite{4DMRI:marklLAA} \\
        \cite{E/A:korea} \\
        \cite{4DMRI:marklLAA}
    \end{tabular}  \\
\hline
\end{tabular}
\smallskip
\caption{\textcolor{black}{Comparison between in silico results and literature data for the four patients in terms of the following biomarkers: velocity magnitudes in LA, at the entrance of the pulmonary veins, at the mitral orifice and at the LAA ostium.}}
\label{tab:peakvelocities}
\end{table}
\par
In Table~\ref{tab:peakvelocities}, we report a validation comparing the peak and mean velocities within the entire LA with medical estimates from MRI images in \cite{4DMRI:garcia,4DMRI:markl}. In addition, we report the LAA emptying peak velocity, which is a widely used indicator because it correlates with thrombus formation. 

%\begin{figure}[t]
	%
%	\centering
	%
%	{\includegraphics[width=0.95\textwidth]{Velocities4.png}}
%	\caption{Volume rendering of phase-averaged velocity magnitude $|\langle\boldsymbol{u}\rangle|_2$ during the A-wave in all the patients.}
%	\label{fig:velfield4}
	%
%\end{figure}
\par
\begin{figure}[t]
	\centering
	{\includegraphics[width=1\textwidth]{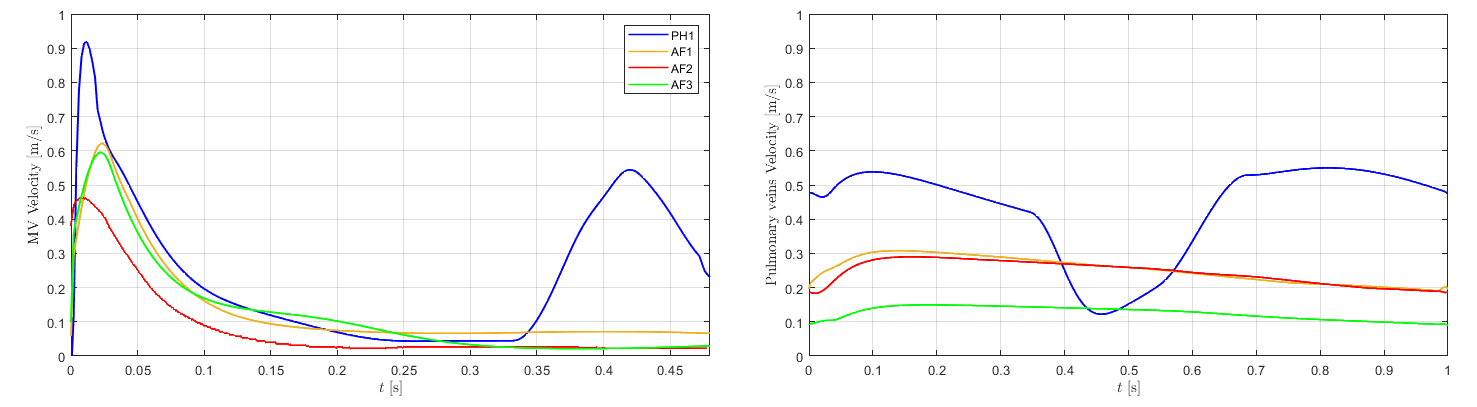}}
	\caption{MV transmitral velocity profiles (left) and pulmonary veins velocity profiles (right) in the four patients.}
	\label{fig:transmitral}
\end{figure}

We report the velocities computed at the MV section in Figure~\ref{fig:transmitral} and we compare our results with the literature estimates from Doppler imaging \cite{E/A:korea,E/A:phys,E/A}. The value is computed by space-averaging the velocity values inside a spherical volume between the leaflet of MV, coherently with the procedure used starting from Doppler images. The case PH1 also allows to perform a validation with the A-wave, not present in AF, peak velocity and then with the E/A ratio, which is an important medical parameter \cite{E/A,E/A:phys}. Finally, we report the mean velocities between the four pulmonary veins computed at the LA entrance in Figure~\ref{fig:transmitral}.

\subsubsection*{Flow stasis}
\begin{figure}[t]
	\centering
	{\includegraphics[width=0.95\textwidth]{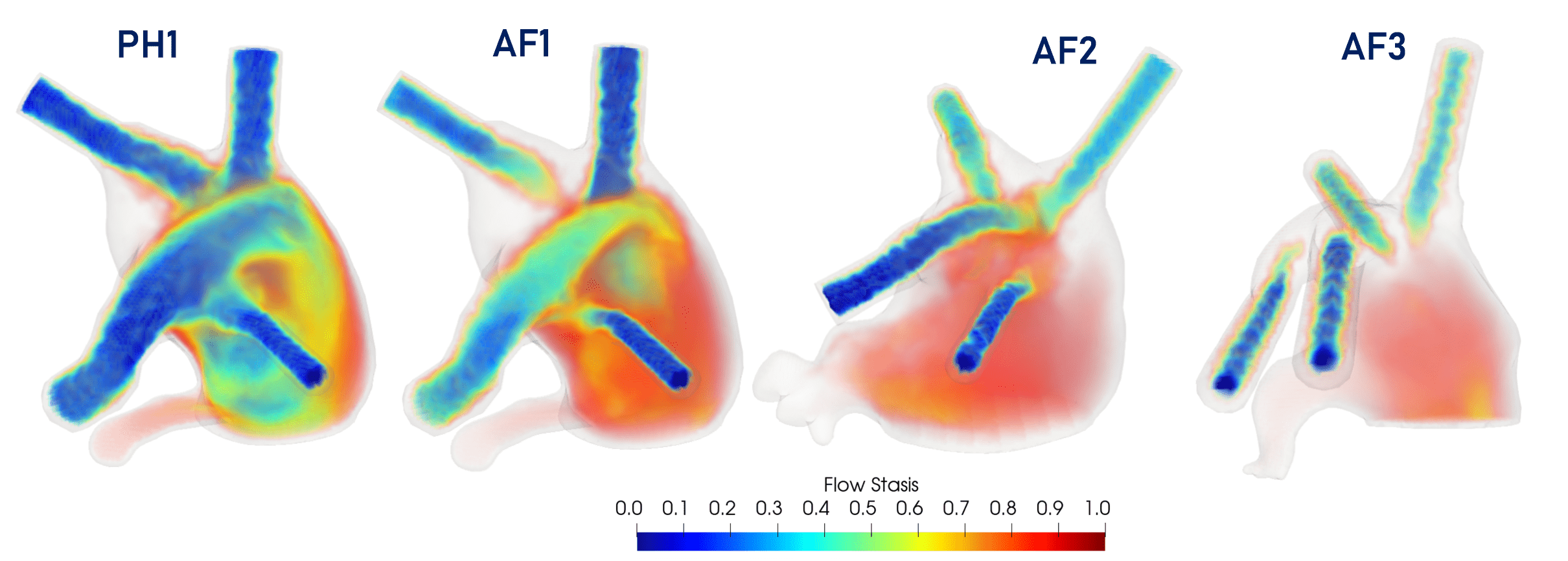}}
	\caption{\textcolor{black}{Volume rendering of the flow stasis indicator $\boldsymbol{\mathrm{F}_\mathrm{S}}$ in the atrial chamber, for all the four cases.}}
	\label{fig:FS}
\end{figure}

The flow stasis is an indicator representing the fraction of time of the heartbeat in which the velocity magnitude in a specific point is smaller than $0.1\; \mathrm{m/s}$ \cite{4DMRI:marklLAA}. This threshold value is consistent with the sensitivity analysis performed in \cite{4DMRI:stasis}. From this indication, we define it as follows:
\begin{equation}
	\mathrm{F_S}(\boldsymbol{\hat{x}}) = \dfrac{1}{T_\mathrm{HB}}\int_0^{T_\mathrm{HB}} \chi_{\{|\langle \boldsymbol{u} \rangle (\boldsymbol{\hat{x}},t)|_2 < 0.1 \mathrm{m/s}\}}(\boldsymbol{\hat{x}},t) \mathrm{d}t \qquad \mathrm{in}\;\widehat{\Omega},
\end{equation}
where $\chi$ is a characteristic function. An accurate estimate of flow stasis is fundamental to determine thromboembolic risk, coherently with the Virchow's triad \cite{virchow:triad}. The volume rendering of the flow stasis for the four patients can be observed in Figure~\ref{fig:FS}. 
\par
However, because of the lower quality of the imaging compared to the CFD resolution it is not possible to make a detailed comparison with the magnitude of the clinical measurements. For this reason, we calculate a spatial-averaged value of flow stasis as in \cite{4DMRI:stasis}. The average is computed neglecting from the atrial domain a boundary layer of $3\;\mathrm{mm}$ width. We report the final values in Table~\ref{tab:flowstasis}.
%To overcome this inconvenience, we will also propose a new indicator in Section \ref{sec:agestasis}, coupling the stasis information with the age of the blood.

\begin{table}[t]
\centering
\begin{tabular}{|c|c|c|c|c|}
\hline
\multirow{2}{*}{\textbf{Patient}}
& \textbf{\textit{In silico}}    & \textbf{Clinical} & \multirow{2}{*}{\textbf{Reference}} \\ 
& \textbf{result} & \textbf{measurement} & \\
\hline
\textbf{PH1}    & $0.333$   
    &   $0.128\div0.502$     & \cite{4DMRI:stasis,4DMRI:marklLAA}  \\
\hline
\textbf{AF1}    & $0.806$   
    &   $0.222\div0.896$     & \cite{4DMRI:stasis,4DMRI:marklLAA}  \\
\hline
\textbf{AF2}    & $0.853$   
    &   $0.222\div0.896$     & \cite{4DMRI:stasis,4DMRI:marklLAA}  \\
\hline
\textbf{AF3}    &$0.873$   
    &   $0.222\div0.896$     & \cite{4DMRI:stasis,4DMRI:marklLAA}  \\
\hline    
\end{tabular}
\smallskip
\caption{ \textcolor{black}{Comparison between in silico results and data coming from literature for the four patients in terms of space-averaged flow stasis.}}
\label{tab:flowstasis}
\end{table}
\subsubsection*{Time-averaged wall shear stress}
\begin{figure}[t]
	\centering
	{\includegraphics[width=0.95\textwidth]{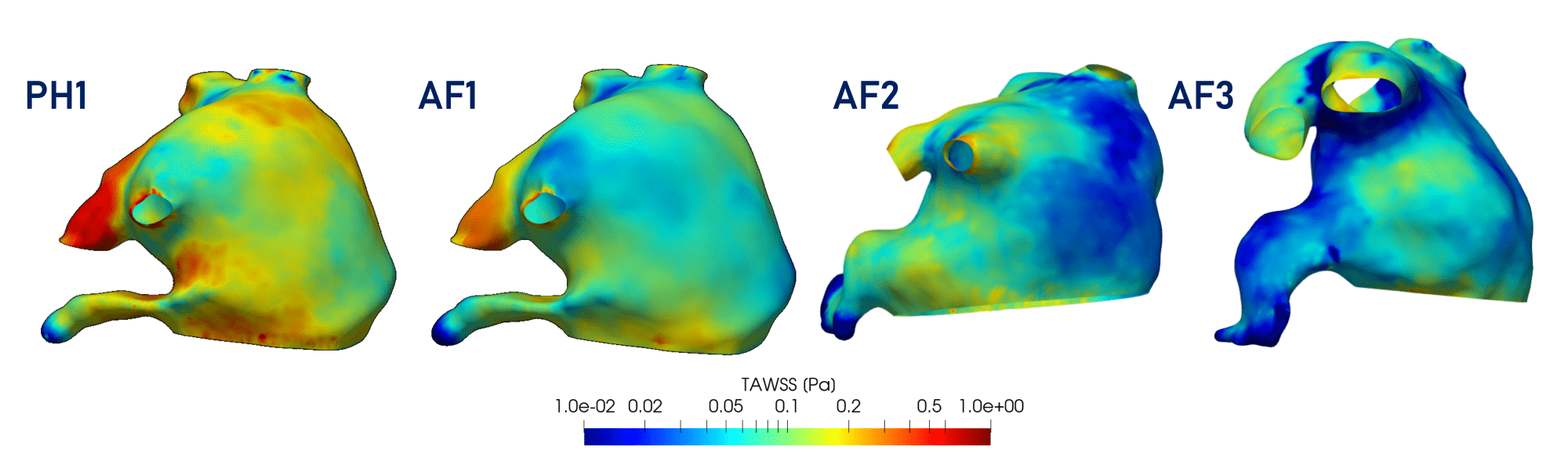}}
	\caption{\textcolor{black}{TAWSS indicator computed on the endocardial surface of atrial chamber for all the four patients.}}
	\label{fig:TAWSS}
\end{figure}

The shear stress at the wall is related to endothelial shear, formation of new tissues and plaques, and promoting of neointimal hyperplasia \cite{WSS:ku}.
% We can define the wall shear stress (WSS) vector on the boundary $\partial\widehat{\Omega}$ as follows:
%\begin{equation}
%	{\boldsymbol{\tau}}_W(\langle\boldsymbol{u}\rangle) = \boldsymbol{\tau}(\langle\boldsymbol{u}\rangle) \cdot \boldsymbol{n} - (\boldsymbol{\tau}(\langle\boldsymbol{u}\rangle) \cdot \boldsymbol{n})\boldsymbol{n} \qquad \mathrm{on}\;\partial\widehat{\Omega},
%\end{equation}
%where $\boldsymbol{\tau}(\langle\boldsymbol{u}\rangle) = 2 \mu \mathbf{D}(\langle\boldsymbol{u}\rangle)$ is the viscous stress tensor and $\mathbf{D}(\langle\boldsymbol{u}\rangle) = \dfrac{1}{2}(\nabla\langle\boldsymbol{u}\rangle+\nabla^\top\langle\boldsymbol{u}\rangle)$.
%We compute the parameter on the reference configuration $\partial\widehat{\Omega}$ provided by the medical images, with the perspective of computing a time-averaged indicator. 
%\par
Specifically, we consider the time-averaged wall shear stress (TAWSS) defined as \cite{koizumi:2014}:
\begin{equation}
	\mathrm{TAWSS}(\langle\boldsymbol{u}\rangle) = \dfrac{1}{T_\mathrm{HB}}\int_0^{T_\mathrm{HB}} |\boldsymbol{\tau}_W(\langle\boldsymbol{u}\rangle)|_2 \,\mathrm{d}t\qquad \mathrm{on}\;\partial\widehat{\Omega},
\end{equation}
where $\boldsymbol{\tau}_W(\langle\boldsymbol{u}\rangle)$ is the wall shear stress (WSS).
In Figure~\ref{fig:TAWSS}, we report the TAWSS for all patients. 

\subsubsection*{Endothelial Cell Activation Potential}
\begin{figure}[t]
	\centering
	{\includegraphics[width=0.8\textwidth]{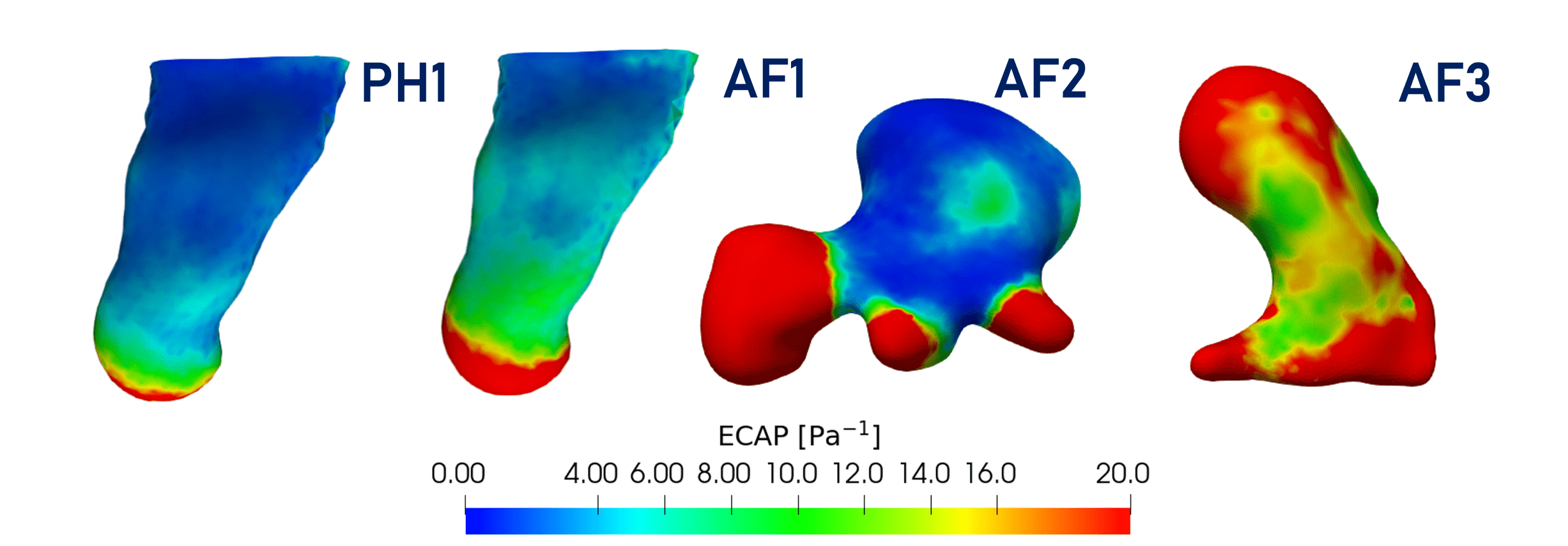}}
	\caption{\textcolor{black}{ECAP indicator computed on the endocardial surface for all the four patients. For each patient, the visualization is restricted only to the LAA.}}
	\label{fig:ECAPLAA}
\end{figure}

Another commonly used indicator to detect the endothelial susceptibility and the thrombus formation probability is the Endothelial Cell Activation Potential (ECAP)  indicator \cite{ECAP:diachille}, defined as:
\begin{equation}
	\mathrm{ECAP}(\langle\widehat{\boldsymbol{u}}\rangle) = \dfrac{\mathrm{OSI}(\langle\widehat{\boldsymbol{u}}\rangle)}{\mathrm{TAWSS}(\langle\widehat{\boldsymbol{u}}\rangle)},
\end{equation}
OSI being the oscillatory shear index \cite{WSS:ku}. OSI is high in regions where WSS changes much during the heart cycle. Thus, ECAP detects high oscillatory and low shear stress regions. In Figure~\ref{fig:ECAPLAA}, we report the ECAP values focus on the LAA wall. 
%\begin{figure}[t]
	%
%	\centering
	%
%	{\includegraphics[width=0.95\textwidth]{ECAP.png}}
%	\caption{ECAP indicator in all the patients.}
%	\label{fig:ECAP}
	%
%\end{figure}
%\par
%From Figure~\ref{fig:ECAP}, we can notice sensibly larger ECAP values in the AF case. This indicator shows a peak for patient AF2, located in between the two left pulmonary veins. However, probably these values are due to the choice to apply no displacement at the veins. Then, we do not find any medical meaning for this local maximum.

\subsection{CFD Lagrangian indicators}
\label{sec:lag}
In the following, we analyze the Lagrangian indicators with the simulations carry out as explained in Section~\ref{sec:lag-explanation}.

\subsubsection*{Mean age of blood}
We consider the injection of particles during all simulated cardiac cycles to construct some Lagrangian fields from the Lagrangian perspective, with the information of all simulated heartbeats. The computation of results based on the age of the blood gives us a new perspective to analyse the regions in which the particles remain for a long time in the atrium, increasing the thrombosis. Additional details on the definition of this field can be found in the Supplementary Material.
\par
The mean age field of red blood cells detects regions in which the blood particles stagnate in the atrium. This indicator was proposed by \cite{pamplona:age} to analyse the age of the blood in LA subjected to AF. We denote by $m_1$ and its definition is provided in the Supplementary Material. In Figure~\ref{fig:m1}, we can observe the $m_1$ field for the four simulated cases. 
\begin{figure}[t]
	\centering
	{\includegraphics[width=0.95\textwidth]{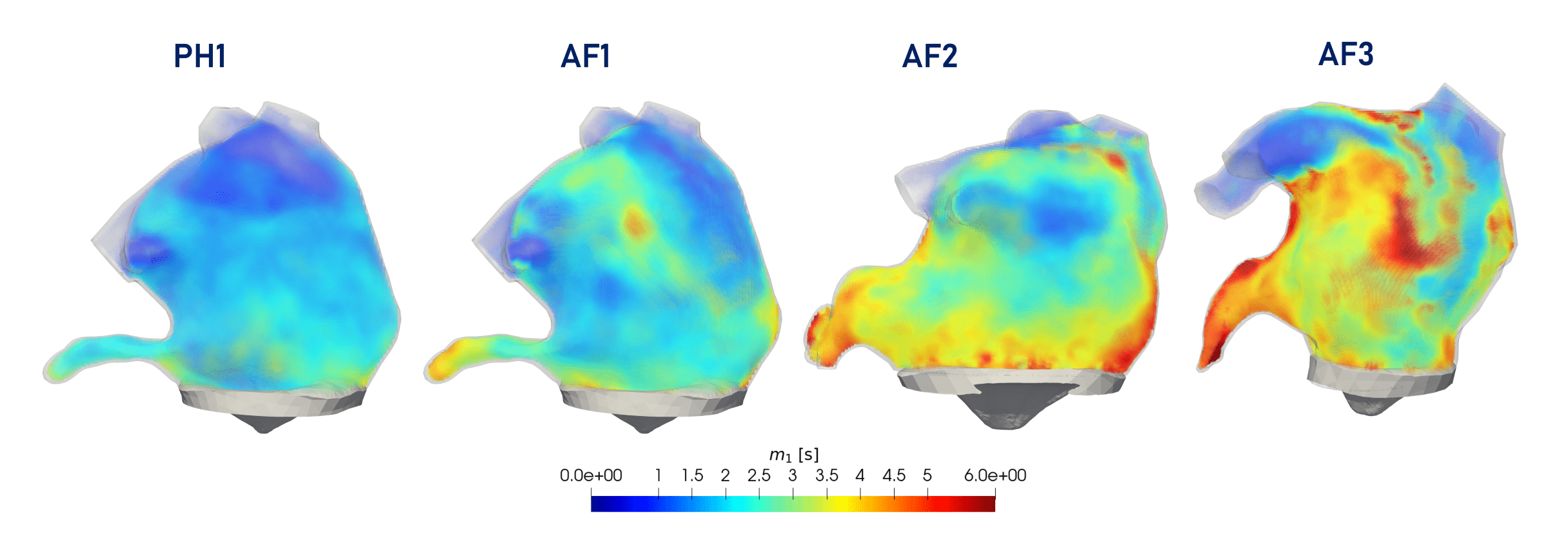}}
	\caption[Blood mean age field]{\textcolor{black}{Volume rendering of the blood mean age field $m_1$ (in seconds) in the atrial chamber for all the four patients.}}
	\label{fig:m1}
\end{figure}

\subsubsection*{Washout of blood}
We compute the washout field $\psi_{\bar{t}}$ at the final time, as defined in the supplementary material. In particular, in the simulation we choose $\bar{t} = 2 \,\mathrm{s}$, we consider a contribution to the field equal to 1 from the particles injected in the first two cycles and equal to 0 for the others. The resulting field gives values between 0 and 1, where:
\begin{itemize}
	\item $\psi_{\bar{t}} \gtrsim 0$, means that there is a local prevalence of particles injected after $\bar{t}$;
	\item $\psi_{\bar{t}} \lesssim 1$, means that there is a local prevalence of particles injected before $\bar{t}$.
\end{itemize}
%Intermediate values are achieved in regions where we do not have a prominent prevalence.
\par
\begin{figure}[t]
	\centering
	{\includegraphics[width=0.9\textwidth]{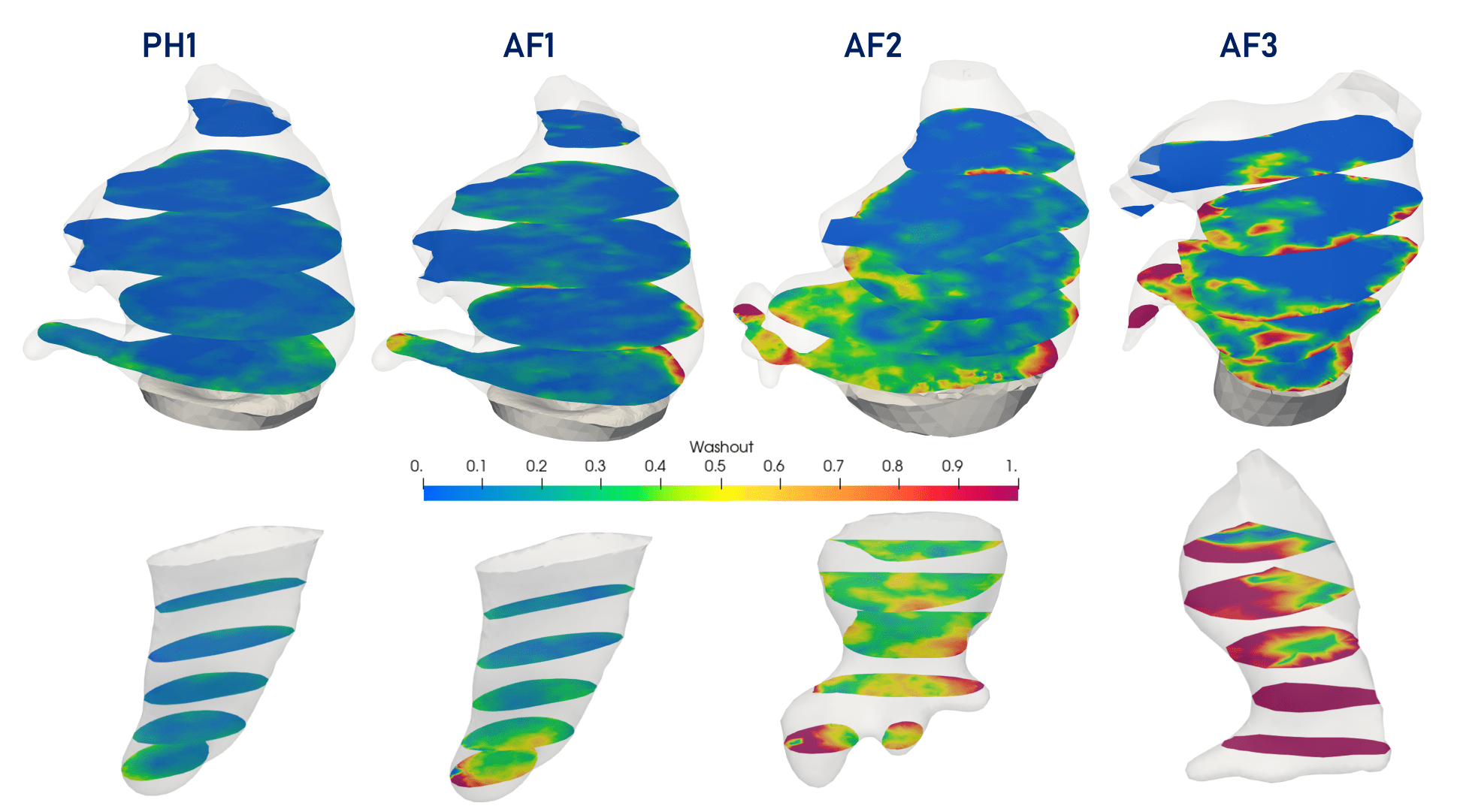}}
	\caption{\textcolor{black}{Washout field $\psi_2$ represented over some slices both in the atrial chambers (first row) and in the LAA (second row) for all the four patients at time $t = 6 \, \mathrm{s}$.}}
	\label{fig:wash}
\end{figure}

In Figure~\ref{fig:wash}, we show the washout field at time $t = 6 \, \mathrm{s}$ using some slices in the LA volume and in its appendage. 
\subsubsection*{Residence time and total path length}
\begin{figure}[t]
	\centering
	{\includegraphics[width=0.9\textwidth]{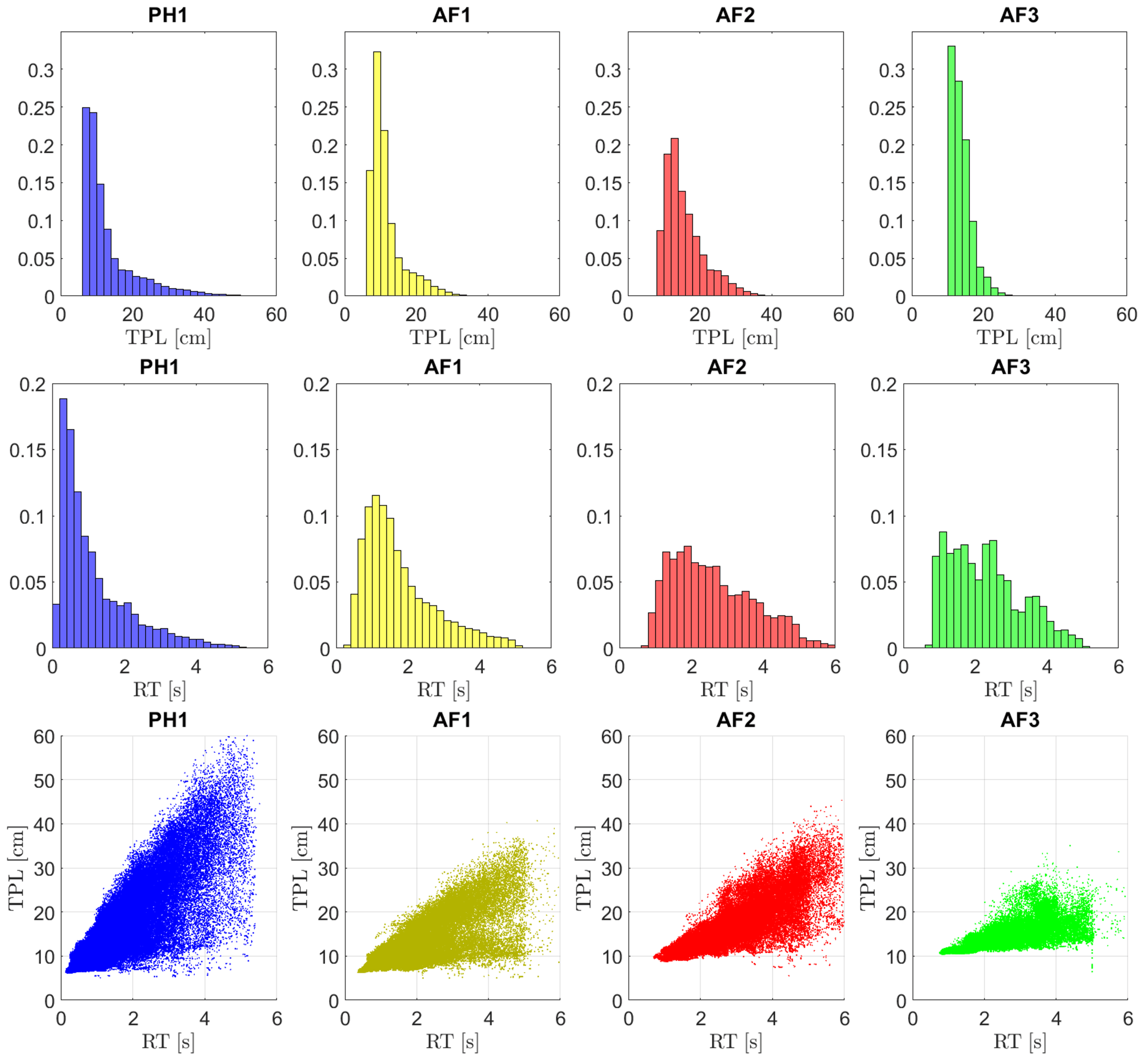}}
	\caption{\textcolor{black}{TPL histogram distribution (first row), RT histogram distribution (second row) and correlation scatter plot (third row) for all the four patients.}}
	\label{fig:RTTPL}
\end{figure}

In this section, we consider particle injection during the first heartbeat only. We analyse the distributions of some Lagrangian indicators in the following five heartbeats of simulation. A video of the motion of the parcels used in this section is provided as supplementary material. In particular, we consider the distribution of two quantities: the Total Path Length (TPL) is the length travelled by a parcel before leaving the LA; the Residence Time (RT) is the time spent by a particle inside the LA. We report the indicators distributions and a scatter plot correlating RT and TPL in Figure~\ref{fig:RTTPL}. 

\subsection{Age Stasis: a new haemodynamic indicator}
\label{sec:agestasis}
\begin{figure}[t]
	\centering
	{\includegraphics[width=0.8\textwidth]{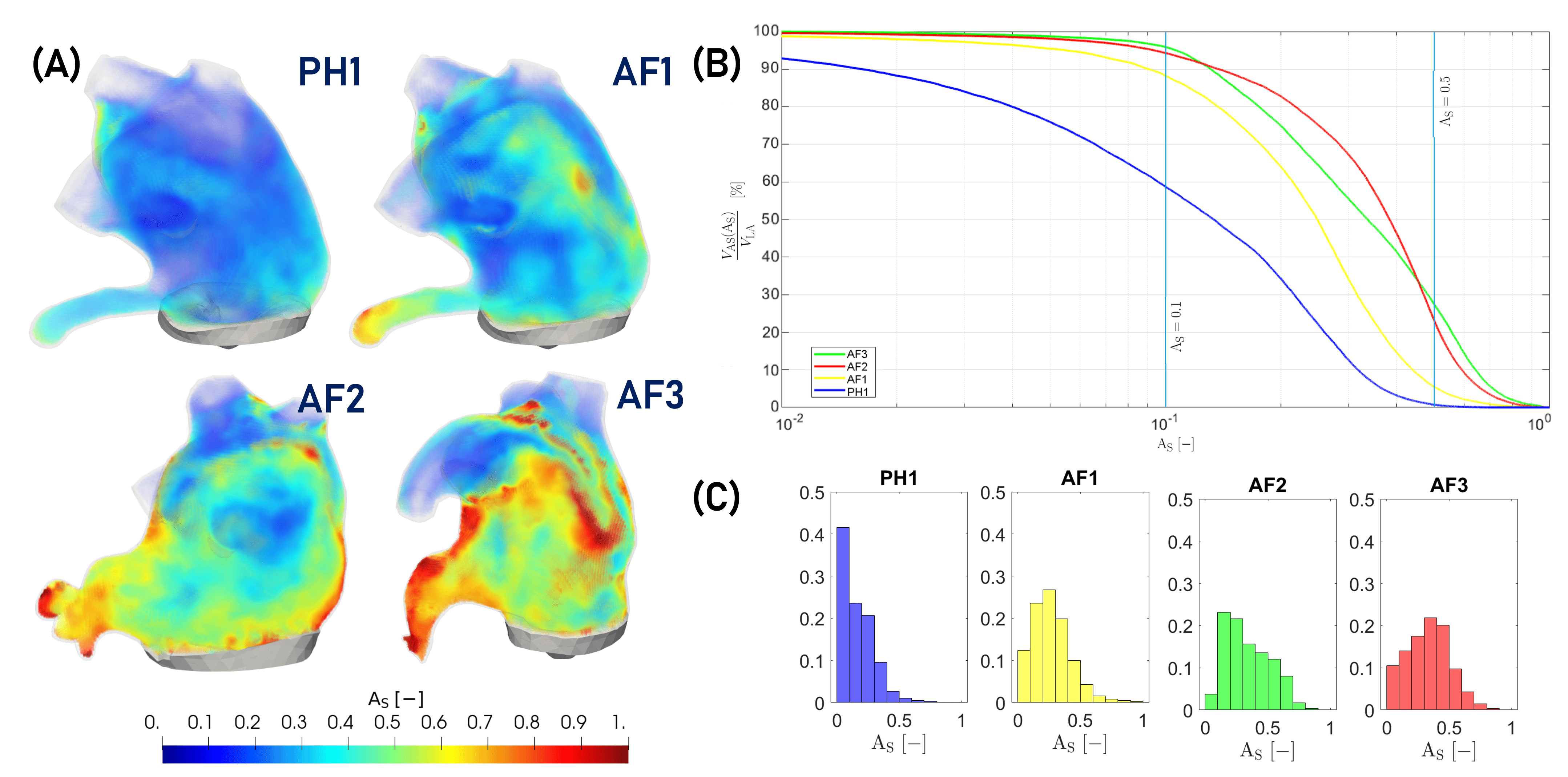}}
	\caption[Age Stasis Volume functions comparison]{\textcolor{black}{Volume rendering of Age Stasis inside the LA (A), Age Stasis Volume function plotted against the Age Stasis values ($x$-axis in logaritmic scale) (B) and histograms of the Age Stasis distributions functions in the LA (C). All the indicators are reported for all the four patients.}}
	\label{fig:agestasis}
\end{figure}

In order to summarise the high complexity of blood flow within LA, we introduce a novel haemodynamic indicator, which we call Age Stasis (AS), by combining results from both Eulerian and Lagrangian analyses of haemodynamics in LA. Our goal is to construct an indicator which considers, at the same time, information about the stagnation of the flow and the oldness of the blood cells. We define AS as:
\begin{equation}
	\mathrm{A_S}(\boldsymbol{x}) = 	\mathrm{F_S}(\boldsymbol{x}) \dfrac{m_1(\boldsymbol{x})}{T}.
\end{equation}
In this way, we are constructing an indicator which accounts for the flow stasis $\mathrm{F_S}(\boldsymbol{x})$, that localizes regions in which the blood flows slowly, and a relative mean age of the blood $m_1(\boldsymbol{x})/T$, being $T$ the final time of the simulation. Thus, we can distinguish regions in which the flow is slow and the particles present are old $(\mathrm{A_S}\simeq 1)$. In order to detect dangerous regions, we neglect $(\mathrm{A_S}\simeq 0)$ the ones in which:
\begin{itemize}
	\item the flow is slow $(\mathrm{F_S}\simeq 1)$, but the particles are not old $(m_1\ll T)$; therefore, there are no particles stationing at that point for a long time. In particular, the boundary layers, where flow stasis is high, are not always associated with a high thrombosis risk;
	\item  the flow is fast $(\mathrm{F_S}\simeq 0)$, but the particles are old $(m_1\simeq T)$. These regions are the ones that require more time to be reached by particles because of their location, but are not stationing points. These regions are not at risk of thrombi formation, because of the absence of the stasis factor. The mitral orifice is the best example of this scenario.
\end{itemize}
The result is a useful dimensionless indicator which assumes values between 0 and 1. We observe in Figure~\ref{fig:agestasis}A the indicator computed in the four simulated cases.

\par

In order to provide a more synthetic quantification of the thrombogenic risk, we introduce the Age Stasis Volume function as:
\begin{equation}
	V_\mathrm{AS}(\theta) = \int_{\widehat{\Omega}} \chi_{\{\mathrm{A_S}(\boldsymbol{x})\geq \theta\}} \mathrm{d}\boldsymbol{x},
\end{equation}
This function, given a value $\theta$ of AS, returns the volume $V_\mathrm{AS}(\theta)$ of blood within the LA characterised by a value of AS greater than $\theta$. Thus, $V_\mathrm{AS}(\theta)$ is useful for investigating the percentage of LA volume associated with a high thrombogenic risk. A comparison between the Age Stasis Volume functions computed in the four simulated cases is provided in Figure~\ref{fig:agestasis}B. By choosing specific values of $\mathrm{A_S}$ we obtain the results in Table~\ref{tab:agestasis}. Finally, in Figure~\ref{fig:agestasis}C, we show the distribution of AS values in the LA volume, constructed using a random sampling of the domain.

\par
\begin{table}[t]
	\centering
	\begin{tabular}{|c|r|r|r|r|}
		\hline
		\textbf{Parameter} & \textbf{PH1} &
		\textbf{AF1} & \textbf{AF2} & \textbf{AF3} \\ 
		\hline
		$\Big(V_\mathrm{LA}-V_{\mathrm{AS}}(0.1)\Big)/V_\mathrm{LA}$
		& $41.3 \%$ & $11.7 \%$ & $5.70 \%$ & $4.20 \% $  \\
		\hline
		$V_{\mathrm{AS}}(0.5)/V_\mathrm{LA}$
		& $0.7 \%$ & $5.5 \%$ & $23.3 \%$ & $27.5 \%$  \\
		\hline
		\end{tabular}
	\smallskip
	\caption{Age Stasis Volumes (\%) significant values.}
	\label{tab:agestasis}
\end{table}

%\begin{figure}[t]
	%
%	\centering
	%
%	{\includegraphics[width=0.8\textwidth]{AgeStasisHist}}
%	\caption[Age Stasis Histogram Distributions]{Age Stasis distribution: histogram visualisation.}
%	\label{fig:agestasishist}
	%
%\end{figure}

\section{Discussion}
\label{sec:disc}
\textcolor{black}{In the following, we discuss the numerical results we obtain. Particularly, we analyze the results of the lumped parameter model in Section~\ref{sec:discussion-0d}. The discussion of the Eulerian and Lagrangian indicators is provided in Sections~\ref{sec:discussion-eulerian} and \ref{sec:discussion-lagrangian}, respectively. Finally Section~\ref{sec:discussion-agesstasis} is devoted to the discussion of the new hemodynamic indicator.}
\subsection{Discussion of lumped-parameter model and displacement}
\label{sec:discussion-0d}
As we can see in Figure~\ref{fig:LPMBCs}, we have two primary changes in pressures distribution. We detect a smaller maximum value of the left ventricular pressure, which is a realistic pathological consequence of AF \cite{pappa:LV,dodge:LV}. The second change is an increase in pulmonary venous pressure, which is also a common effect of the pathology; these values can be found in advanced pathologies \cite{PVH}, as simulated ones. In the AF case, we have less variation in the pulmonary venous flow rate; in particular, we have lower values due to reduced LAEF. Indeed, AF cases do not show the ``atrial kick'' due to the reduced contractility deriving from the incorrect action potential diffusion  \cite{iwasaki:AF,schotten:AF}.
\par
\textcolor{black}{The LA contraction can be observed in detail in the video provided in the supplementary material. The displacement of the wall cannot be directly validated due to the lack of in vivo recordings. However, we can observe some similarities by comparing the displacement with others provided by electromechanical simulations in literature \cite{suppvideomotion,whole:electro}.
The level of contraction of both atrium and LAA visible in \cite{suppvideomotion} is coherent with what we imposed. Concerning the LAA motion, level of contraction seems to be homogeneous on the surface; this result is analogous to what we obtain by adding the component directed to the LAA centre of mass. Both similarities derive from the correction term of the analytical formulation we propose. Nevertheless, a deeper analysis of the calibration of the parameters could improve our displacement accuracy.}
\par
\textcolor{black}{Maintaining fixed pulmonary vein entrances seems to be a good approximation of what is shown by these simulations \cite{suppvideomotion, whole:electro}. A fixed MV orifice represents a limitation of our study; the motion of the mitral annulus is evident in electromechanical simulations of the whole heart \cite{whole:electro}. However, we can overcome this problem within the parametric displacement setting as done in \cite{zingaro:multiscale}. The second limitation is that the LAA contraction lags slightly behind the atrial chamber due to the action potential diffusion \cite{whole:electro}. This phenomenon is not reproduced by our model.}
\subsection{Discussion of CFD Eulerian indicators}
\label{sec:discussion-eulerian}
All the values we compute starting from our simulations are consistent with the ranges present in the medical literature (Table~\ref{tab:peakvelocities}). The values of LAA emptying peak velocity are slower in AF conditions, consistently with the literature \cite{4DMRI:marklLAA}. The inlet velocity at the pulmonary vein is significantly lower in patient AF3. This might be related to the geometrical differences between patients; as a matter of fact, AF3 is characterized by a smaller atrial chamber and larger veins. The peak velocities at the MV section detected both under physiological and AF conditions are consistent with medical estimates, as shown in Table~\ref{tab:peakvelocities}. Also the parameters of the A-wave of PH1 are consistent with those present in the medical literature.
\par
The values of the curves in Figure~\ref{fig:transmitral} are coherent with those estimates available in literature \cite{4DMRI:garcia,4DMRI:marklLAA}. In particular, we validate peak velocities during the heartbeat at the pulmonary vein inlet compared to estimates from MRI images \cite{4DMRI:garcia,4DMRI:marklLAA}, which are reported in Table~\ref{tab:peakvelocities}. Concerning the two relative maxima reached by the curve of the physiological case, we found that they have approximately the same magnitude. This property is typical for young patients without any cardiovascular disease \cite{demarchi:pvs}. The lack of information about peak velocities at pulmonary veins in persistent AF does not allow the validation of the remaining two simulations in terms of these indicators.
\par
From the analysis of the flow stasis indicator in Figure~\ref{fig:FS}, we note larger stasis values under AF conditions, consistent with the slower velocities we found. Moreover, clinical measurements provided by the 4D flow MRI data in literature confirm this trend \cite{4DMRI:marklLAA,4DMRI:spartera}. The averaged quantities in Table~\ref{tab:flowstasis} are within the reference ranges. The exclusion of the boundary layer is fundamental; otherwise, considering a global average would cause an overestimate of the stasis due to the no-slip condition at the wall. 
\par
The analysis of TAWSS in Figure~\ref{fig:TAWSS} and ECAP in Figure~\ref{fig:ECAPLAA} confirms that in AF cases, we obtain smaller values than in physiologic conditions. This is significant, in particular considering the same geometry by comparing PH1 and AF1. This is associated with a significantly higher risk of thrombosis in patients with AF. In general, we achieve the minimum values on the final part of the LAA surface, consistently with the medical literature \cite{LAA:al-saady, LAA:ernst, LAA:tan}. However, in AF2, the morphology of the atrium allows larger values in the first part of the appendage, which is large and allows the formation of some vortices, causing a better washout of the LAA\footnote{A detailed analysis of the vorticity in LA can be found in \cite{Corti:Tesi}.}. On the contrary, the morphology of the appendage of patient AF3 causes a slower blood flux due to its flat morphology. The indicators confirm the location of high thrombosis risk  inside the LAA \cite{LAA:al-saady,LAA:ernst,LAA:tan}. These results are coherent with the ones obtained from other patient-specific studies present in literature \cite{ECAP:LAA,ECAP:LAA2}, correlating the LAA geometries to the ECAP spatial distribution.
\par
Eulerian analysis clearly shows the higher risk of thrombosis formation in AF and located in LAA, considering different types of indicators, connected to both two parameters of the Virchow's triad: blood stasis (flow stasis, velocity magnitude and endothelial susceptibility (TAWSS and ECAP). However, the limit of the Eulerian analysis is the absence of historical notions about the blood flow, connected with the particles path, and only detectable  using a Lagrangian perspective.

\subsection{CFD Lagrangian indicators}
\label{sec:discussion-lagrangian}
In Figure~\ref{fig:m1}, we observe that regions with the largest values of mean blood age are near the boundary of the MV orifice and in the LAA. In conditions of AF, the regions where the mean age field is greater than or equal to $3\,\mathrm{s}$ are much more than in the physiological case. Differences are evident, in particular, considering persistent fibrillation in patients AF2 and AF3, probably due to higher volumes, lower velocities, and lower ejection fractions.  
\par
At the same time, in patients affected by AF, the washout field shows incomplete blood turnover in LAA and in the lower part of the atrial chamber (Figure~\ref{fig:wash}). However, distinct geometries show some differences: patient AF3 has more static blood than AF2 inside the appendage and near the boundaries. While in the middle of the atrium, we find better washout values that are consistent with the larger LAEF. Regarding LAA, the surfaces clearly show the differences between PH1 and AF1. In fact, in the apical part of LAA, we have less blood exchange under fibrillation conditions, but we cannot detect any difference near the ostium.
\par
In the end, in Figure~\ref{fig:RTTPL}, we can notice that particles that remain in the atrium for more than a heartbeat under AF conditions follow shorter paths on average. This fact is consistent with the lower velocities detected in the blood, which cause shorter instantaneous steps. Furthermore, in the PH1 case, we have more particles that leave the LA faster, consistent with the larger LAEF found in Table~\ref{tab:patientdata}. 
\par 
Finally, we observe that not all particles that remain for several heartbeats in the chamber cover paths of comparable length. As a matter of fact, within the particles characterized by the same RT value, some particles follow long paths, whereas others cover smaller ones. The latter makes the blood more stagnant, causing the formation of blood clots. The difference between these two types of paths suggests that the analysis of the mean age field cannot provide a complete analysis of thrombosis risk; to complement the indicator, we couple it with the flow stasis to obtain a more comprehensive indicator in Section \ref{sec:agestasis}.

\subsection{Age Stasis: a new haemodynamic indicator}
\label{sec:discussion-agesstasis}
AS assumes low values in the whole domain in the PH1 case; in contrast, the indicator assumes large values under AF conditions in several parts of the geometry (Figure~\ref{fig:agestasis}A). The regions with maximum stasis are located near the LAA. These results are coherent with the discussion in the previous sections.
\par
In Figure~\ref{fig:agestasis}B, the cumulative distributions show that the percentage of volume associated to low values of stasis is significantly higher in physiological than in AF conditions, for each value of AS. We can observe the same trend by making a comparison between paroxysmal and permanent AF cases. For this reason, we can infer a relation among the resulting curve and the severity of the pathology. The results in Table~\ref{tab:agestasis} show that, in PH1, more than the $40\%$ of the total volume is associated to values of AS smaller than this threshold. On the contrary, in AF, we have a sensible reduction in volume. Furthermore, for AF2 and AF3, there are lower values ($5-6\%$) than in the AF1 case ($\simeq10\%$), consistent with the persistence of AF in those cases.
\par
Another interesting result is the evaluation of the volume percentage, which presents a value of AS greater than $0.5$. In this way, we quantify the regions in which we have a higher thrombogenic risk. The results in Table~\ref{tab:agestasis} confirm that there is a higher risk in fibrillation cases. In particular, in advanced pathologies, the volume percentage associated with this risk is approximately a quarter of the total; on the contrary, under physiological conditions, we detect a percentage smaller than $1\%$. In Figure~\ref{fig:agestasis}C, the distribution of PH1 case is different from the others due to the predominancy of low stasis values. AF2 and AF3 show distribution tails that are more consistent in the region with high stasis values than AF1.
\section{Conclusions}
\label{sec:conclusions}
In this article, we numerically simulated left atrial haemodynamics under physiological and atrial fibrillation conditions, considering patient-specific geometries and parametric analytical displacement fields. We examined the numerical results from a Eulerian and Lagrangian point of view, computing several indicators and biomarkers used in the literature. Finally, we propose a novel haemodynamic indicator to analyse the risk of thrombosis by combining the two approaches; moreover, we use it to compute a synthetic distribution function, which allows a quick comparison between different individuals.
\par
We introduced an original procedure to compute pressure, flow rates and a parametric displacement field: they serve as boundary conditions for our CFD problem. Specifically, we employ a lumped-parameter 0D circulation model, and we tune it to simulate the haemodynamics in either physiological or pathological conditions, being respectful of the geometrical constraints given by the patient-specific atria (i.e. the maximum left and right atrial volumes). We use a ``one-way'' 0D-3D coupling scheme between the circulation and the 3D CFD problem. We introduce a new parametric displacement field that correctly catches the typical ejection fraction values of the left atria and their auricles. Using the proposed procedure, we can set up the numerical simulation under physiological and pathological conditions.

The results detected a substantial reduction of blood velocity and shear stresses on the endocardial walls in pathological conditions. AF increases the average time that a single particle spends in the left atrium and reduces the washout of the chamber. In addition, we found that the variability among patients in terms of morphological features of the left atrial appendage impacts its haemodynamics. We found that a large ostium improves the LAA washout. Additionally, the possible existence of lobes in the apical part can significantly affect the blood dynamics.

%The Eulerian analysis detected a substantial reduction of blood velocity and shear stresses on the endocardial walls in pathological conditions. Lagrangian simulations confirmed that AF increases the average time that a single particle spends in the left atrium and reduces the washout of the chamber. In addition, we found that the variability among patients in terms of morphological features of the left atrial appendage impacts its haemodynamics. We found that a large ostium improves the LAA washout. Additionally, the possible existence of lobes in the apical part can significantly affect the blood dynamics.

Furthermore, by coupling Eulerian and Lagrangian results, we proposed a novel hemodynamic indicator, Age Stasis, that accurately detects regions associated with a high risk of thrombosis, by searching, at the same time, for slow flow conditions and the presence of old blood. The Age Stasis allows us to highlight the regions where blood clots formation is most probable. Furthermore, the cumulative distribution functions provided the ability to make comparisons between different patients, quantifying how advanced pathology influences the risk of thrombosis, within a single, synthetic indicator.

\subsection{Limitations and Future Developments}
%Several future developments are possible to overcome the limitations of this work. First, it can be interesting to apply our procedure to determine boundary data (pressure, flowrates, and displacement fields) to patients for whom we can also detect all this information from medical images. In this way, we can perform validation to assess the accuracy level of the novel procedure, in particular concerning the wall motion. Moreover, if 4D flow MRI are available, the discrepancy of the numerical results from 3D time-dependent velocity data from medical images can be investigated.
%\par
%Another development consists of the evaluation of the indicators performing a CFD simulation in CT-derived geometries. These would be less smooth, and a higher geometrical complexity could impact the haemodynamics, especially in the LAA. Furthermore, in order to assess the ventricular flow in case of AF, one could carry out CFD simulation of the whole left heart. This study could be functional in evaluating the effects of fibrillation on ventricular flow patterns.
\textcolor{black}{The absence of a patient-specific validation of the procedure is the main limitation of this work. Moreover, the lack of medical images limits the calibration of the models (lumped-parameter model, wall displacement) which required to make many assumptions. It can be interesting to apply our procedure to determine boundary data (pressure, flowrates, and displacement fields) to patients for whom we can also detect all this information from medical images. In this way, we can perform validation of the atrial wall motion (in particular in the LAA region) and, if 4D flow MRI is available, of the CFD numerical results.}
\par
%\textcolor{black}{Another limitation of this study is the use of MRI-derived geometries. The use of CT-derived ones, which would be less smooth, and with a higher geometrical complexity could impact the haemodynamics, especially in the LAA.} 
\textcolor{black}{Another limitation of this study is the use of MRI-derived atrial chambers in contrast to computed tomography, which is known to produce less smooth geometries. Indeed, a higher geometrical complexity may influence the atrial haemodynamics, especially in the LAA.}
Furthermore, in order to assess the ventricular flow in case of AF, one could carry out a CFD simulation of the whole left heart. This study could be functional in evaluating the effects of fibrillation also on  ventricular flow patterns.
\par
Finally, due to the electric nature of fibrillatory arrhythmia, a comparison between the indicators we discussed in this paper with some electromechanical ones  could be interesting.  In this direction, physics-based atrial models would allow us to validate our boundary displacement procedure and, at the same time, to better highlight the limitations of our novel approach.

\section*{Declaration of competing interests}
The authors declare that they have no known competing financial interests or personal relationships that could have appeared to influence the work reported in this article.

\section*{Acknowledgments}
%This work is based on Data from the Zenodo Database "Constructing a Human Atrial Fibre Atlas" (https://doi.org/10.5281/zenodo.3764917) open accessible under conditions of Creative Commons Attribution 4.0 International Public License.

AZ, LD and AQ have been funded by the Italian Ministry of University and Research (MIUR) within the PRIN  2017 project «Modeling the heart across the scales: from cardiac cells to the whole organ» Grant Registration number 2017AXL54F). The authors acknowledge the anonymous Reviewers for their insightful comments and suggestions.

\appendix

\section{Parameter values of lumped-parameter model}

We report the values chosen for the parameters of the lumped-parameter circulation model in Table \ref{tab:0Dparam}.

\begin{table}[t]
 \begin{subtable}[h]{0.45\textwidth}
	\centering
	\begin{tabular}{|c|r l|}
		\hline
		\textbf{Parameter} & \multicolumn{2}{c|}{\textbf{Value}} \\ 
		\hline 
		$R_\mathrm{AR}^\mathrm{SYS}$
		& $1.00$ & $\mathrm{mmHg\cdot s/mL}$  \\ 
		\hline 
		$C_\mathrm{AR}^\mathrm{SYS}$
		& $2.00$ & $\mathrm{mL/mmHg}$  \\ 
		\hline 
		$L_\mathrm{AR}^\mathrm{SYS}$
		& $1.00$ & $\mathrm{mmHg\cdot s^2/mL}$  \\ 
		\hline 
		\hline 
		$R_\mathrm{VEN}^\mathrm{SYS}$
		& $0.50$ & $\mathrm{mmHg\cdot s/mL}$  \\ 
		\hline 
		$C_\mathrm{VEN}^\mathrm{SYS}$
		& $140.00$ & $\mathrm{mL/mmHg}$  \\ 
		\hline 
		$L_\mathrm{VEN}^\mathrm{SYS}$
		& $5\times 10^{-4}$ & $\mathrm{mmHg\cdot s^2/mL}$  \\ 
		\hline 
		\hline 
		$R_\mathrm{AR}^\mathrm{PUL}$
		& $0.04$ & $\mathrm{mmHg\cdot s/mL}$  \\ 
		\hline 
		$C_\mathrm{AR}^\mathrm{PUL}$
		& $15.00$ & $\mathrm{mL/mmHg}$  \\ 
		\hline 
		$L_\mathrm{AR}^\mathrm{PUL}$
		& $5\times10^{-4}$ & $\mathrm{mmHg\cdot s^2/mL}$  \\ 
		\hline 
		\hline 
		$R_\mathrm{VEN}^\mathrm{PUL}$
		& $0.60$ & $\mathrm{mmHg\cdot s/mL}$  \\ 
		\hline 
		$C_\mathrm{VEN}^\mathrm{PUL}$
		& $10.00$ & $\mathrm{mL/mmHg}$  \\ 
		\hline 
		$L_\mathrm{VEN}^\mathrm{PUL}$
		& $1\times 10^{-4}$ & $\mathrm{mmHg\cdot s^2/mL}$  \\ 
		\hline 
	\end{tabular}
	
	\smallskip
	\caption{Parameter values used in the four simulations.}
	\label{tab:common0Dparam}
	\end{subtable}
    \begin{subtable}[h]{0.45\textwidth}
	\centering
	\begin{tabular}{|c|r|r|r|r|}
		\hline
		\textbf{Parameter} & \textbf{PH1} &
		\textbf{AF1} & \textbf{AF2} & \textbf{AF3} \\ 
		\hline 
		$E_\mathrm{RA}^\mathrm{act,max}$
		& $0.04$ & $0.00$ & $0.00$ & $0.00$  \\ 
		\hline  
		$E_\mathrm{RA}^\mathrm{pass}$
		& $0.06$ & $0.60$ & $0.60$ & $0.80$  \\ 
		\hline
		\hline
		$E_\mathrm{RV}^\mathrm{act,max}$
		& $1.20$ & $1.20$ 
		& $0.70$ & $0.70$  \\ 
		\hline  
		$E_\mathrm{RV}^\mathrm{pass}$
		& $0.05$ & $0.80$ 
		& $0.40$ & $0.40$  \\ 
		\hline 
		\hline
		$E_\mathrm{LA}^\mathrm{act,max}$
		& $0.20$ & $0.00$ 
		& $0.00$ & $0.00$  \\ 
		\hline  
		$E_\mathrm{LA}^\mathrm{pass}$
		& $0.09$ & $0.30$ 
		& $0.30$ & $0.40$  \\ 
		\hline
		\hline
		$E_\mathrm{LV}^\mathrm{act,max}$
		& $6.00$ & $4.00$ 
		& $1.00$ & $1.00$  \\ 
		\hline  
		$E_\mathrm{LV}^\mathrm{pass}$
		& $0.08$ & $0.20$ 
		& $0.20$ & $0.20$  \\ 
		\hline
	\end{tabular}
	\smallskip
	\caption{Elastances values (in $\mathrm{mmHg/mL}$).}
	\label{tab:0Dela}
	\end{subtable}
	\caption{Parameter values of the lumped-parameter models for the four patients.}
	\label{tab:0Dparam}
\end{table}

\printcredits

\bibliographystyle{model1-num-names}
%\bibliographystyle{cas-model2-names}

\bibliography{cas-refs}



\end{document}

\let\WriteBookmarks\relax
\def\floatpagepagefraction{1}
\def\textpagefraction{.001}

\shorttitle{Kinetic models for red blood cells}    

\shortauthors{M. Corti, A. Zingaro, L. Dede', A. Quarteroni}  

\title [mode = title]{Supplementary material: Kinetic models for red blood cells}  

%\tnotemark[<tnote number>] 

%\tnotetext[<tnote number>]{<tnote text>} 


\author[Polimi]{Mattia Corti}

\cormark[1]


\ead{mattia.corti@polimi.it}

%\ead[url]{<URL>}

%\credit{<Credit authorship details>}

\affiliation[Polimi]
{organization={MOX-Dipartimento di Matematica, Politecnico di Milano},%Department and Organization
            addressline={Piazza Leonardo da Vinci 32}, 
            city={Milan},
            postcode={20133},
            country={Italy}}
            
\author[Polimi]{Alberto Zingaro}
%\ead{alberto.zingaro@polimi.it}
\author[Polimi]{Luca Dede'}
%\ead{luca.dede@polimi.it}
\author[Polimi,EPFL]{Alfio Maria Quarteroni}
%\ead{alfio.quarteroni@polimi.it}

\affiliation[EPFL]{organization={Chair of Modelling and Scientific Computing (CMCS), Institute of Mathematics, \'{E}cole Polytechnique F\'{e}d\'{e}rale de Lausanne},%Department and Organization
            addressline={Station 8, Av. Piccard}, 
            city={Lausanne},
            postcode={CH-1015}, 
            country={Switzerland (Professor Emeritus)}}

\maketitle

A Lagrangian approach is helpful and it is frequently used in heart haemodynamics to analyse the red blood cells motion inside LA  \cite{zingaro:LA,chnafa:LES,pamplona:age}. The application of a kinetic theory approach allows the derivation from the particle information of some Eulerian fields, as fields of moments of the fluid age \cite{sierra:age, spalding:age} and the washout field \cite{pamplona:age}.
\par
The evaluation of the moments starting from an Eulerian perspective would require an additional computational cost because it would need, at each timestep, the resolution of a specific transport-diffusion PDE \cite{pamplona:age}. Moreover, the LES framework gives us the possibility to simulate the transport and diffusion of the particles using the Navier-Stokes velocity solution without the requirement of any assumption about the diffusion tensor, whose modeling would sensibly affect the final result. For these reasons, in this section we want to provide a brief derivation of the moments of the age and washout fields, used in the paper. In \ref{sub:MSP} we analyse the single-particle motion to support the typical assumption of considering the red blood cell as a massless particle \cite{chnafa:LES} with a dimensional analysis. In \ref{sec:Boltz}, we introduce the kinetic theory to derive the evolution PDEs in \cite{sierra:age,pamplona:age}, from our perspective. Finally, in \ref{sec:kinnum}, we present the discretization procedure.

\section{Equation of motions for a single particle in a fluid}
\label{sub:MSP}
The Lagrangian approach models the blood as a fluid represented by the plasma, filled with particles transported by the flow \cite{vergara:actanumerica}. In particular, we focus on the red blood cells because of their high concentration (about the $45$\% is made of red blood cells and the $55$\% of plasma and the remaining particles occupy less than $1$\% of the total volume).
\par
Considering the cells as single particles, for which we can neglect both the gravitational effects and the collisional interaction between themselves, the equation of motion of the $P$-th one can be written as follows:
\begin{equation}
	\dfrac{\mathrm{d}\boldsymbol{v}_P(t)}{\mathrm{d}t} = \dfrac{1}{\tau_P}(\boldsymbol{u}(\boldsymbol{x}_P(t),t)-\boldsymbol{v}_P(t))\qquad t\in(0,T),
	\label{eq:vel}
\end{equation}
where $\tau_P$ is the particle response time in our flow regime \cite{zaichik:particle,burton:particles}, $\boldsymbol{u}$ is the fluid velocity field, and $\boldsymbol{x}_P$ and $\boldsymbol{v}_P$ are the particle position and velocity, respectively. Sure of the correctness of Equation \ref{eq:vel}, we want to make some simplifications to reduce the problem to the resolution of an interpolation problem.
\par
Under the Stokes flow assumption, we would have a response time $\tau_0$ with the analytical formulation \cite{zaichik:particle}:
\begin{equation}
	\tau_0 = \dfrac{\rho_P d^2_P}{18\mu},
\end{equation}
where $\mu$ is the fluid dynamic viscosity, $d_P = 7.8 \times 10^{-6}\; \mathrm{m}$ is the red blood cell diameter and $\rho_P = 1.11 \mathrm{g/mL}$ is the blood cell density. In our case we need a correction of particle response time, due to the transitional regime of the flow, which becomes the following \cite{zaichik:particle}:
\begin{equation}
    \label{eq:taup}
	\tau_P = \dfrac{\tau_0}{\phi(\mathrm{Re}_P)},
\end{equation}
where $\mathrm{Re}_P$ is the particle Reynolds number \cite{zaichik:particle}, given by: 
\begin{equation}
	\mathrm{Re}_P = \dfrac{d_P|\boldsymbol{u}(\boldsymbol{x}_P(t),t)-\boldsymbol{v}_P(t)|_2\rho}{\mu}.
\end{equation}
If we compute the $\mathrm{Re}_P$, using an estimate of $|\boldsymbol{u}(\boldsymbol{x}_P(t),t)-\boldsymbol{v}_P(t)|_2\simeq 10^{-4}\;\mathrm{m/s}$ \cite{Sugii:2005}, we obtain $\mathrm{Re}_P\simeq 3\cdot10^{-4}$.
\par
In Equation \ref{eq:taup}, we use the Schiller-Neumann approximation (see \cite{clift:particles}) for $\mathrm{Re}_P<10^3$:
\begin{equation}
	\phi(\mathrm{Re}_P) = 1 + 0.15\mathrm{Re}_P^{0.687},
\end{equation} 
then we obtain $\phi(\mathrm{Re}_P)\simeq1$. For this reason, we have the possibility to make the approximation $\tau_P\simeq\tau_0$. We can define the Stokes number to compare it with the flow time scale $\tau$ as \cite{zaichik:particle}:
\begin{equation}
    \label{eq:stokes}
	\mathrm{St} = \dfrac{\tau_{0}}{\tau} = \dfrac{\rho_P d_P^2 U}{\mu L},
\end{equation}
where and $U$ and $L$ are the characteristic velocity and length, respectively. 
\par
Substituting the values in Equation \ref{eq:stokes}, we can notice that $\mathrm{St}\ll 1$. This value justifies the approximation of the red cells with a tracer, as in \cite{chnafa:LES}. Then the particle velocity can be approximated with the value of the fluid velocity in the particle position:
\begin{equation}
	\boldsymbol{v}_P(t) = \boldsymbol{u}(\boldsymbol{x}_P(t),t).
\end{equation}
Finally, the equation of motion of the $P-$th single red blood cell:
\begin{equation}
	\begin{cases}
		\dfrac{\mathrm{d}\boldsymbol{x}_P(t)}{\mathrm{d}t} = \boldsymbol{v}_P(t) \\
		\boldsymbol{v}_P(t) = \boldsymbol{u}(\boldsymbol{x}_P(t),t)
	\end{cases}
	\label{eq:pm}
\end{equation}
\section{Kinetic equation for the particle distribution}
\label{sec:Boltz}
The description of a large number of particles requires the introduction of a statistical approach. First of all, we consider $N_P$ particles describing the $j$-th one, using the position $\boldsymbol{x}_j(t)$ and velocity $\boldsymbol{v}_j(t)$. Starting from that the microscopic distribution function in the phase space can be defined as the product of two Dirac delta functions $\delta$:
\begin{equation}
	f_m(\boldsymbol{x},\boldsymbol{v},t) = \sum_{j=1}^{N_P} \delta(\boldsymbol{x}-\boldsymbol{x}_j(t))
	\delta(\boldsymbol{v}-\boldsymbol{v}_j(t)),
\end{equation}
where, the first $\delta$ is a function of the spatial variable, and the second one is velocity-dependent.
The distribution $f_m$ evolves according to Liouville equation for the particle transport \cite{zaichik:particle}:
\begin{equation}
	\dfrac{\partial f_m}{\partial t}+\boldsymbol{v}\cdot\dfrac{\partial f_m}{\partial \boldsymbol{x}}+\dfrac{\partial}{\partial \boldsymbol{v}}\cdot\Big(\dfrac{\boldsymbol{u}-\boldsymbol{v}}{\tau_P}f_m\Big) = 0.
\end{equation}
\par
Then we apply an ensemble average operation $\langle\cdot\rangle$ over the random realizations of the velocity field $\boldsymbol{u}$ to separate the fluctuating components of the turbulent motion (see \cite{zaichik:particle}), as follows:
\begin{equation}
	f_m = \langle f_m\rangle + \tilde{f}_m = f + \tilde{f}_m
	\qquad \boldsymbol{u} = \langle \boldsymbol{u}\rangle + \tilde{\boldsymbol{u}} = \boldsymbol{U} + \tilde{\boldsymbol{u}}
\end{equation}
This provides the definition of $f(\boldsymbol{x},\boldsymbol{v},t)$ which is the probability density for a particle to have position $\boldsymbol{x}$ and velocity $\boldsymbol{v}$ at instant time $t$. The function $f$ evolves according to the following equation \cite{zaichik:particle}:
\begin{equation}
	\dfrac{\partial f}{\partial t}+\boldsymbol{v}\cdot\dfrac{\partial f}{\partial \boldsymbol{x}}+\dfrac{\partial}{\partial \boldsymbol{v}}\cdot\Big(\dfrac{\boldsymbol{U}-\boldsymbol{v}}{\tau_P}f\Big) = -\dfrac{1}{\tau_P}\dfrac{\partial}{\partial \boldsymbol{v}}\cdot\Big<\boldsymbol{\tilde{u}}\tilde{f}_m\Big>.
	\label{eq:bolt}
\end{equation}
\par
This equation can be used to detect the Eulerian equations which describe the problem from a continuous mechanics point of view. Indeed, we can define:
\begin{itemize}
	\item the average volume concentration:
	\begin{equation}
		\Phi(\boldsymbol{x},t)=\int_{\mathbb{R}^3} f(\boldsymbol{x},\boldsymbol{v},t) \mathrm{d}\boldsymbol{v}, 
	\end{equation}
	\item the disperse phase's velocity:
	\begin{equation}
		\boldsymbol{V}(\boldsymbol{x},t) = \dfrac{1}{\Phi(\boldsymbol{x},t)} \int_{\mathbb{R}^3} \boldsymbol{v} f(\boldsymbol{x},\boldsymbol{v},t) \mathrm{d}\boldsymbol{v},
	\end{equation}
	which allows us to construct a decomposition of the particles velocity in $\boldsymbol{v} = \boldsymbol{V} + \boldsymbol{v}'$, where $\boldsymbol{v}'$ is the fluctuating component \cite{zaichik:particle}.
\end{itemize}
\par
In our case, following the procedure in \cite{zaichik:particle} and considering $\tau_P\rightarrow 0$ since $\mathrm{St}\ll 1$, we can derive the equations:
\begin{equation}
	\begin{cases}
		\boldsymbol{V} = \boldsymbol{U} - \mathbf{D} \cdot \nabla \ln(\Phi) & \mathrm{in}\;\mathbb{R}^3,\\
		\dfrac{\partial \Phi}{\partial t} + \nabla\cdot (\Phi\boldsymbol{U}) = \nabla\cdot (\mathbf{D}\cdot\nabla\Phi) & \mathrm{in}\;\mathbb{R}^3,
	\end{cases}
\label{eq:tracer}
\end{equation}
where $\mathbf{D}$ is a tensor which describes the turbulent diffusion of the particles \cite{zaichik:particle}.
In our context, we have inflow and outflow of blood. For this reason, we define a particle distribution function $f_\tau$ depending on the injection time of the particle $\tau$ in the atrium. 
\par
Then the average volume concentration of the tracer injected at the time $\tau$ is:
\begin{equation}
	\Phi_\tau(\boldsymbol{x},t,\tau) = \int_{\mathbb{R}^3} f_\tau(\boldsymbol{x},\boldsymbol{v},t,\tau) \mathrm{d}\boldsymbol{v}.
\end{equation}
Following the steps we described before considering the dependence on $\tau$ we arrive to:
\begin{equation}
	\rho_P\dfrac{\partial \Phi_\tau}{\partial t} + \nabla\cdot (\rho_P\Phi_\tau\boldsymbol{U}) = \nabla\cdot (\rho_P\mathbf{D}\cdot\nabla\Phi_\tau)
	\label{eq:taueq}
\end{equation}
which is the evolution equation $\tau$-family of \cite{sierra:age} for a non reactive tracer, with the only difference that $\Phi_\tau$ is not exactly a distribution because its integral in space gives us the number of particles and not $1$.
\par 
In this case, we need some minor changes in the formulation of the moment generations, using:
\begin{equation}
	\Phi(\boldsymbol{x},t) = \int_{-\infty}^{t} \Phi_\tau(\boldsymbol{x},t,\tau)\,\mathrm{d}\boldsymbol{x},
\end{equation}
which represents the average volume concentration at time $t$, considering all the particles injected until that moment. This can be used in the definition of the $k$-th moment of the age:
\begin{equation}
	m_k(\boldsymbol{x},t) =  \dfrac{1}{\Phi(\boldsymbol{x},t)}\int_{-\infty}^{t}(t-\tau)^k \Phi_\tau(\boldsymbol{x},t,\tau) \mathrm{d}\tau \qquad k\in\mathbb{N},
\end{equation}
which can be related with an integration over $\tau$ of the equation \ref{eq:taueq} to the following evolution equation of the $k$-th moment:
\begin{equation}
	\dfrac{\partial (\rho_P\Phi m_k)}{\partial t} + \nabla\cdot (\rho_P\Phi m_k \boldsymbol{U}) = \nabla\cdot (\rho_P\mathbf{D}\cdot\nabla(\Phi m_k)) + \rho_P\Phi k m_{k-1},
	\label{eq:momenteq}
\end{equation}
This is the commonly used moment generation equation \cite{spalding:age,sierra:age,pamplona:age}. It is interesting to notice that a common assumption made using Equation \ref{eq:momenteq} is $m_0=1$. From our procedure, this condition is automatically derived by the kinetic definition of $m_k$.
\par
The idea of the separation of the function $f$ using the continuous family of functions $(f)_{\tau\in\mathbb{R}}$ gives us the possibility to derive the moment generation from the kinetic theoretical background arriving to a theoretically equivalent result.
\par
Another indicator we want to connect to our Lagrangian formulation is the washout field, which can be defined as:
\begin{equation}
	\psi_{\bar{t}}(\boldsymbol{x},t) =  \dfrac{1}{\Phi(\boldsymbol{x},t)}\int_{-\infty}^t H(\bar{t}-\tau)\Phi_\tau(\boldsymbol{x},t,\tau) \mathrm{d} \tau
\end{equation}
where $H$ is the Heaviside function.
This parameter gives us a concentration $\mathrm{Im}(\psi_{\bar{t}}) = [0,1]$ of the tracer injected after the time $\bar{t}$ on the overall tracer concentration. In particular, when $\psi_{\bar{t}} = 0$, we have high concentration of young tracer \cite{pamplona:age}. Indeed, if we consider $t<\bar{t}$ then in the integration $\tau<t$ for each $\tau$ and $\psi_{\bar{t}} = 1$.
\par
Then, if we multiply the equation \ref{eq:taueq} by the Heaviside function and we integrate over $\tau$ we have the equation:
\begin{equation}
	\dfrac{\partial (\rho_P\Phi\psi_{\bar{t}})}{\partial t} + \nabla\cdot (\rho_P\Phi\psi_{\bar{t}}\boldsymbol{U}) = \nabla\cdot (\mathbf{D}\cdot\nabla(\rho_P\Phi\psi_{\bar{t}}))
\end{equation}
which is the scalar transport equation used to describe the washout.
\section{Numerical approximation}
\label{sec:kinnum}
The particle simulation can be segregated from the NS problem, indeed we assumed that the particles do not effect the fluid motion. The simulation will be so carried out using an approximation of the particle velocity and position as follows for $n = 0,..., N_{\theta}-1$:
\begin{equation}
	\begin{cases}
		\boldsymbol{x}_P^{n+1} = \boldsymbol{x}_P^{n} + \Delta \theta \boldsymbol{v}_P^{n} 	\\
		\boldsymbol{v}_P^{n} = \boldsymbol{u}(\boldsymbol{x}_P^n)
\end{cases}
\end{equation}
where $\Delta \theta = T/N_\theta$ is the time discretization step, which can be also bigger than the one we use to solve the Navier-Stokes problem, if compatible with the stability requirements of an Explicit Euler method. This approach needs an interpolation of the velocity field we computed in an Eulerian way. Indeed, the $\boldsymbol{u}$ values are stored at the mesh grid points, which do not match with the particle position at any time.
\par
The velocity numerical approximation will be the following \cite{bazilevs:VMS}:
\begin{equation}
	\boldsymbol{u} = \boldsymbol{u}_h + \boldsymbol{u}' \simeq \boldsymbol{u}_h - \tau_M(\boldsymbol{u}_h)\boldsymbol{r}_M(\boldsymbol{u}_h,p_h).
\end{equation}
Indeed, choosing the FEM solution $\boldsymbol{u}_h$ and neglecting the turbulence component of the velocity $\boldsymbol{u}'$, the approximation on coarse meshes yields poor results than those of DNS \cite{chibbaro:parcels}, causing infinite residence times of particles near to the boundary, in analogy with the negligible turbulent diffusion in the PDEs \cite{liu:age}.
\par
The choice of the fine-scale component consideration in the tracking is commonly used in Lagrangian Parcels Tracking in LES framework \cite{yama:parcels,chibbaro:parcels}. However, no results are present in literature about a Lagrangian analysis, coupled with a VMS-LES framework and in most cases, the fine-scale corrections use stochastics Brownian motion to model $\boldsymbol{u}'$ \cite{chibbaro:parcels,huilier:parcels}. 
\par
The approach we will use to simulate the red blood cells constructs some macro-particles, which we will call from now on parcels. This approximation, known as Discrete Parcel Method (DPM), is commonly used in many different applications \cite{DPM}. Moreover, the Particle-In-Cell Method for plasma simulations \cite{PIC} or the Moist-Parcel-In-Cell Method for cloud particles \cite{MPIC} use the same idea, and they are commonly used to recover Eulerian fields. We consider $N_H$ parcels, and we define a weight for the simulation, which is the fraction between the number of physical particles over the number of simulated particles:
\begin{equation}
	\omega_j = \dfrac{N_P}{N_H}.
\end{equation}
We consider a constant $\omega = \omega_j,\; \forall j=1,...,N_H$. Moreover, we assume that a single parcel $j$ has a definite velocity, but also a spatial extension, given by a shape function $S$. In this way, we can approximate the distribution function as follows:
\begin{equation}
f_h(\boldsymbol{x},\boldsymbol{v},t) = \sum_{j=1}^{N_H} \omega_j\,S(\boldsymbol{x}-\boldsymbol{x}_j(t))\,\delta(\boldsymbol{v}-\boldsymbol{v}_j(t)).
\end{equation}
Our choice about the function $S$ will be a 3D extension of the cloud-in-cell described by \cite{inpro:lapenta}:
\begin{equation}
	S(|\boldsymbol{x}|) = \dfrac{1}{4\pi R^3/3}\,\chi_{\{|\boldsymbol{x}|<R\}},
\end{equation}
where $R = R(h)$ is a function of the mesh element size; we will use it to compute the weight of a particle. 
The choice of this value needs to respect some physical constraint. First of all, we cannot consider a sphere too small to contain $\omega_j$ particles, because we cannot overestimate the density of cells:
\begin{equation}
	R \geq R_{\omega_j} = \sqrt[3]{\dfrac{3V_{\omega_j}}{4\pi}},
\end{equation} 
where $V_{\omega_j}$ is the minimum volume which contains a number $\omega_j$ of red blood cells, considering that the mean estimate is about $5.5\times10^6\,\mathrm{cells}/\mathrm{mm}^3$. In our simulations, we choose the following expression of $R$:
\begin{equation}
	R = h = \dfrac{1}{|\mathcal{T}_h|}\sum_{K\in\mathcal{T}_h} h_K,
\end{equation}
where $h$ is the average mesh size.
\par
Then, the function $S$ has the properties of symmetry, compact support and unitary integral, which are fundamental in the derivation of motion equations, which can be proved to be the same of the real single particle, following the procedure of \cite{inpro:lapenta} on our equations. 
\par
The utility of this procedure is the derivation of the Eulerian fields from the Lagrangian one. Indeed, considering each parcel $j$ entering in the domain at time $\Theta_j$, we can compute an approximation of the age $K$-th moment field at each grid node $\boldsymbol{x}_i$, as the age $K$-th moment average over the parcels which are distant from the point less than $R$:
\begin{equation}
	m_{K,h}^n(\boldsymbol{x}_i) = \dfrac{\sum_{j=1}^{N_H} (t-\Theta_j)^K\omega_j\, S(\boldsymbol{x}_i-\boldsymbol{x}_{j}^n)}{\sum_{j=1}^{N_H} \omega_j\,S(\boldsymbol{x}_i-\boldsymbol{x}_j^n)}
\end{equation}
\par
Finally, we can use the same calculation also to find the washout from the same perspective:
\begin{equation}
	\psi_{\overline{t},h}^n(\boldsymbol{x}_i) = \dfrac{\sum_{j=1}^{N_H} H(\overline{t}-\Theta_j)\,\omega_j\,S(\boldsymbol{x}_i-\boldsymbol{x}_{j}^n)}{\sum_{j=1}^{N_H} \omega_j\,S(\boldsymbol{x}_i-\boldsymbol{x}_j^n)}
\end{equation}

\printcredits

\bibliographystyle{model1-num-names}
%\bibliographystyle{cas-model2-names}

\bibliography{cas-refs}

